 [2005/06/22 5.2/AAS markup document class]%
\documentclass[preprint]{aastex}%

\usepackage{natbib}
\bibpunct{(}{)}{,}{a}{}{;}
\usepackage{epsfig}
\usepackage{amssymb}
\usepackage{morefloats}

\newcommand{\g}{\mbox{$\rm\,g$}}

\newcommand{\cm}{\mbox{$\rm\,cm$}}

\newcommand{\myr}{\mbox{$\rm\,Myr$}}
\newcommand{\gyr}{\mbox{$\rm\,Gyr$}}
\newcommand{\yr}{\mbox{$\rm\,yr$}}
\newcommand{\s}{\mbox{$\rm\,s$}}
\newcommand{\hour}{\mbox{$\rm\,h$}}

\newcommand{\km}{\mbox{$\rm\,km$}}
\newcommand{\m}{\mbox{$\rm\,m$}}

\newcommand{\tiunits}{\mbox{$\rm\,J\,m^{-1}\,s^{-1/2}\,K^{-1}$}}

\newcommand{\beq}{\begin{equation}}
\newcommand{\eeq}{\end{equation}}
\newcommand{\po}{\phantom{1}}

\begin{document}



\title{Coupled Spin and Shape Evolution of Small Rubble-Pile Asteroids:
Self-Limitation of the YORP Effect}

\author{Desire\'e Cotto-Figueroa\altaffilmark{1,2}, Thomas S.~Statler\altaffilmark{1,3,4}, Derek C.~Richardson\altaffilmark{4} and Paolo Tanga\altaffilmark{5}}
\email{dcottofi@asu.edu}

\email{statler@ohio.edu}

%
%
%
%
%

\altaffiltext{1}{Astrophysical Institute, Department of Physics and Astronomy 251B Clippinger Research Laboratories, Ohio University, Athens, OH 45701, USA}
\altaffiltext{2}{School of Earth and Space Exploration, Arizona State University Tempe, AZ 85287 USA}
\altaffiltext{3}{Division of Astronomical Sciences, National Science Foundation, 4201 Wilson Blvd, Arlington, VA 22230, USA}
\altaffiltext{4}{Department of Astronomy, University of Maryland, College Park, MD,20742, USA}
\altaffiltext{5}{Laboratoire Lagrange, UMR7293, Universit\'e de Nice Sophia-Antipolis, CNRS, Observatoire de la C\^ote d'Azur, BP 4229, 06304 Nice Cedex 4, France}

\begin{abstract}
We present the first self-consistent simulations of the coupled spin-shape
evolution of small gravitational aggregates under the influence of the YORP
effect. Because of YORP's sensitivity to surface topography, even small
centrifugally driven reconfigurations of aggregates can alter the YORP torque
dramatically, resulting in spin evolution that can differ qualitatively from
the rigid-body prediction. One third of our simulations follow a simple
evolution described as a {\em modified YORP cycle}. Two-thirds exhibit one or
more of three distinct behaviors---{\em stochastic YORP}, {\em self-governed
YORP}, and {\em stagnating YORP\/}---which together result in {\em YORP
self-limitation.} Self-limitation confines rotation rates of evolving
aggregates to far narrower ranges than those expected in the classical YORP
cycle, greatly prolonging the times over which objects can preserve their
sense of rotation. Simulated objects are initially randomly packed, disordered
aggregates of identical spheres in rotating equilibrium, with low internal
angles of friction. Their shape evolution is characterized by rearrangement
of the entire body, including the deep interior. They do not evolve to
axisymmetric top shapes with equatorial ridges. Mass loss occurs in one-third
of the simulations, typically in small amounts from the ends of a
prolate-triaxial body. We conjecture that YORP self-limitation may inhibit
formation of top-shapes, binaries, or both, by restricting the amount of
angular momentum that can be imparted to a  deformable body. Stochastic YORP,
in particular, will affect the evolution of collisional families whose orbits
drift apart under the influence of Yarkovsky forces, in observable ways.

\end{abstract}
 
 



\section{Introduction}

The distribution of asteroids with diameters larger than a few hundred meters in the period-diameter diagram
is interpreted widely as evidence that these
objects are not monolithic boulders \citep{Dav79, Har96}. 
The sharp cutoff in rotation period at $P \approx 2\hour$
matches the spin rate at which material at
the equator of a rocky sphere would become gravitationally
unbound; the persistence of this envelope to large sizes implies that
these objects are dominated by gravity, obscuring the
effects of tensile or shear strength \citep{Hol07}.
Their actual structures may range from contact configurations
of a few monolithic blocks to nearly homogeneous
collections of individual small grains.  Direct measurements of the
masses and volumes of 433 Eros and 25143 Itokawa by the NEAR-Shoemaker
and Hayabusa spacecraft imply porosities of 27\%
\citep{Wil02} and 40\% \citep{Abe06}, respectively,
arguing for both fractured bodies and genuine rubble
piles in the near-Earth asteroid (NEA) population.

In sharp contrast, NEAs smaller than about $150\m$ in diameter
overwhelmingly are rotating faster than the $2\hour$ limit.  These
objects are under centrifugal tension in directions perpendicular to
the spin axis, and under gravitational compression along it.  Despite
an initial rush to dub them ``monolithic fast rotators,'' it was shown
by \citet{Hol07} that geological granular materials can supply
sufficient cohesion to hold aggregate bodies together at the observed
sizes and spin rates.  The most surprising aspect of the fast-rotating
asteroids (FRAs) is their
abrupt appearance as a function of absolute magnitude:
essentially everything smaller than $H=23.6$ (nominal diameter
$\sim 60\m$), and nothing larger than $H=21.4$ ($\sim 170\m$), is a
fast rotator \citep{Sta13}.  This abrupt transition is not
predicted by current strength models \citep{Hol07,San14}.

Owing to the action of the YORP effect---the secular torque due to the
reflection and thermal re-emission of solar radiation from the surface
\citep{Pad69,RuBot00,Rub00,BotReview}---the current
spins of NEAs with diameters ($D$) of a few km or
smaller may not reflect their original
spin states. YORP spin timescales $|P/(dP/dt)|$ in the inner Solar System are 
$\sim 10^6 (D / 1\km)^2 \yr$ \citep{Rub00}, as confirmed by observational
detections of YORP acceleration
\citep{Low07,Tay07,Kas07,Dur08a,Dur08b,Dur12,Low14}.
Typical NEA lifetimes are $\sim 10^7\yr$ \citep{Gla97},
so there is ample opportunity, in principle,
for YORP to modify the spins of sub-km-sized NEAs.

For a given object and orbit, the secular YORP torque is a fixed vector
function of obliquity. It has become standard practice to use the plot of
the torque components {\it vs.} obliquity---the ``YORP curves''\footnote{Or
sometimes ``Rubincam curves.''}---as a description of the YORP characteristics
of an object. If the object remains rigid, the YORP curves determine
its spin evolution: the so-called ``YORP cycle'' \citep{RuBot00}.  A typical
cycle begins with the object at an obliquity at which the torque component
along the spin axis is positive; the object accelerates in spin
rate and evolves in obliquity until it reaches an orientation at which
the spin component changes sign, then decelerates while evolving
toward an end-state obliquity that is a stable fixed point.  Once the
spin period is comparable to the orbital period, spin-orbit resonances
come into play; these, along with tides or small
impacts, randomly re-orient the rotation axis, possibly after an episode
of slow chaotic tumbling, to an obliquity at which the cycle can begin anew.

The YORP cycle concept has important implications for orbital evolution
driven by the Yarkovsky effect (the net radiation recoil force),
which itself is spin-state dependent.
Most NEAs are thought to have been delivered from the Main Belt to their
current orbits with retrograde rotation, having drifted inward (via
Yarkovsky) to various resonances \citep{Bot02, LaS04}. Once in the inner
Solar System, YORP timescales should become short. As the asteroids complete
their YORP cycles, their previous spin states would be forgotten, and the
preference for retrograde rotation should be erased. Yet, recent
observational determinations of Yarkovsky semi-major axis drift rates from
available radar and optical astrometry find that the overwhelming majority
have $da/dt < 0$, indicating retrograde rotation \citep{Che08, Nug12, Far13}. This
is difficult to reconcile with simple time-scale arguments showing
that YORP should have been able to re-write the spin state distribution of
sub-km-sized objects many times over. 

The possibility that the YORP cycle may accelerate objects to high rotation
rates has excited interest in spin-driven reshaping and binary formation,
a compelling demonstration of which is presented by \citet{WalshNature}.
These authors simulate idealized self-gravitating aggregate asteroids
composed of identical spheres, assumed to be inexorably accelerated by YORP.
They find that the objects with a suffciently high internal angle of friction, 
or with a rigid core, become oblate and develop an equatorial ridge, making the 
body resemble a child's top.  Continued spin-up causes the ridge to shed
material, which can then reaccrete in orbit.  This process dynamically
associates binaries with top shapes; and the strong
resemblance of the simulated binary formed by \citet{WalshNature}
to the actual binary 1999 KW$_4$ \citep{Ost06} is striking. YORP is now
widely held to be an important mechanism in binary formation. But
this belief rests on the assumption that YORP will, first,
accelerate objects to spin rates high enough to form axisymmetric tops;
then, accelerate the tops so that they shed mass; and finally, drive
sufficient mass off the surface and into orbit to form a binary companion.
Simulations to date have adopted the {\em ansatz\/} that YORP will provide
angular momentum in whatever amount is needed to accomplish this. But this is
not a safe assumption when the object is not a rigid body.

Deformability, as one would expect for a rubble pile, fractured body,
or anything with loose surface material, may significantly alter the
behavior of the YORP effect.
Because the net YORP torque is a small residual of an imperfect cancellation
of competing contributions across the asymmetric surface, YORP is inherently
sensitive to the internal mass distribution and to the detailed surface
topography. \citet{Sch08} demonstrated that $\sim 50\m$ shifts of
25143 Itokawa's center of mass could change the sign of the spin component
of torque, an effect subsequently confirmed by \citet{Low14}.
\citet{StatlerYORP} systematically studied the topographic
effect on a wide variety of simulated asteroids, and showed that objects that
are identical but
for the location of a single crater or boulder can have torques differing by
factors of several. \citet{StatlerYORP} further conjectured that
the successive effects of minor structural changes that alter the surface
may qualitatively alter spin evolution under YORP, possibly replacing the
YORP cycle with a stochastic random walk at rotation periods
$\lesssim 10\hour$, and potentially limiting the amount of angular momentum
that YORP can contribute to processes like rotational reshaping and binary
formation.

The purpose of this paper is to test the conjecture of \citet{StatlerYORP}
through self-consistent numerical simulations of coupled shape and spin
evolution of gravitationally bound aggregates driven by the YORP effect.
We will demonstrate that {\em stochastic YORP\/} can, indeed, occur, and
is just one of three distinct processes deriving from spin-driven shape
change, that collectively give rise to {\em YORP self-limitation}. Section
\ref{s.methods} describes our numerical approach and the simulated
aggregates that we use for our initial conditions. Section \ref{s.results}
presents the results, describing the time evolution in spin and obliquity
as well as the statistics of mass reconfigrations, shape changes, and
mass shedding; it also presents a preliminary version of a statistical
(Monte Carlo) description of self-consistent spin evolution. Section
\ref{s.discussion} discusses the implications for top shapes, binaries,
and the Yarkovsky effect, and Section \ref{s.summary} sums up.

\section{Numerical Methods}\label{s.methods}

\subsection{Overview}

The physical system we are simulating is characterized by two very different
timescales: the dynamical timescale---$10^3\s$ to $10^5\s$---on
which the object rotates and centrifugally driven material movement may
occur, and the YORP timescale---$10^{13}\s$ to $10^{15}\s$ for
kilometer-sized objects---on which the spin state is altered. Running a
discrete-element simulation for $10^{10}$ dynamical times is not feasible,
but we can exploit the difference in timescales. Material reconfigurations,
quick compared with the YORP timescale, take place at effectively
constant angular momentum; and YORP evolution, acting slowly between
reconfigurations, takes place at constant shape. This allows us to adopt a
two-step computational approach in which we integrate the YORP-induced spin
state evolution at constant shape, incrementing (or decrementing) the spin rate
in the discete-element code on a greatly compressed timescale until material
movement is triggered, and then follow the dynamical evolution in ``real''
time, at constant angular momentum, until the reconfiguration is finished. At
that point we recompute the torques for the new shape and resume the spin state
integration. This back-and-forth approach, handing off between the particle
dynamics and the radiation dynamics parts of the calculation, is the key to
making these simulations possible.

\subsection{Gravity and Particle Dynamics: {\tt pkdgrav}}

The gravitational and particle dynamics are simulated using the hard-sphere
discrete element method (HSDEM) as implemented in {\tt pkdgrav}, a
gravitational $N$-body tree code originally developed for cosmology
\citep{StadelThesis} and subsequently modified to handle interparticle
collisions \citep{Ric00,Ric09,Ric11}. The ensemble of spherical particles
used by {\tt pkdgrav}
is intended to model the collective behavior of a deformable material composed
of discrete pieces, not to literally represent components of the
aggregate. Collisions between pairs of spheres are treated as instantaneous
events that alter their translational and rotational motions. Dissipative
effects are parametrized by coefficients of restitution that affect the
relative motion of the surfaces at the point of contact in the normal
($\varepsilon_n$) and tangential ($\varepsilon_t$) directions. To avoid an
unmanageable number of extremely low-velocity collisions, $\varepsilon_n$
and $\varepsilon_t$ are set to 1 (no dissipation) below a threshhold controlled by two additional
parameters termed the {\em collapse limit\/} and the {\em slide limit.} Because dissipative processes in small asteroids are not quantitatively well understood, we do not attempt at this stage to
model the rate of relaxation or the lifetimes of non-principal-axis (NPA)
rotation states. Cohesive forces can, in principle, be included, but are ignored in the simulations
reported here; hence the results are applicable to objects in the
few-kilometer size range where gravity dominates, and are not easily scalable
to smaller sizes where cohesion is expected to become relatively more important.

Aggregates modeled by the HSDEM approach may be somewhat more deformable than
real aggregates composed of irregularly shaped components, owing to the ability
of the spherical particles to roll. The use of identical
spheres allows for a certain degree of rigidity resulting from ``cannonball
stacking''. \citet{Ric05} and \citet{Wal12} find that cannonball-stacked
arrangements of identical spheres in hexagonal-close-pack (HCP) configuration
have angles of friction near $40\arcdeg$, comparable to lunar and martian
regolith. When the spheres are not in ordered packing, the resulting aggregates
have angles of friction in the range of $5\arcdeg$ to $10\arcdeg$. This
is lower than typical values for terrestrial granular materials;
however, the properties of real asteroidal materials are not quantitatively
well determined.  \citet{Tan09}, using the {\tt pkdgrav} HSDEM
implementation, demonstrate that a population of disordered aggregates of
identical spheres, allowed to equilibrate at constant angular momentum, can
collectively reproduce the observed asteroid shape distribution. On the basis
of this result we adopt the objects from the \citet{Tan09} study as our test
objects in this paper. These choices represent a simple starting point, a first step in
simulations of self-consistent spin evolution.  In Section 4 we describe physical
mechanisms and computational strategies that will be appropriate for subsequent steps.

\subsection{Radiation and Surface Physics: {\tt TACO}}

The dynamical effects of radiation recoil are calculated using {\tt TACO}
\citep{StatlerYORP}, a code for calculating thermophysical processes on the
surfaces of inactive small bodies. {\tt TACO} models an asteroid
surface using a triangular tiling. The interaction of each tile with incident
solar radiation is described by a Hapke model for the bidirectional
reflectance \citep{HapkeV}. Shadowing is handled explicitly by calculating a
horizon map for each tile, which gives the maximum elevation of the visible
parts of the surface as a function of azimuth from the tile centroid. The
incident radiation that is not reflected is absorbed and heats the surface.
{\tt TACO} includes the ability to solve the 1-dimensional heat conduction
equation for the flow of heat into and out of the surface; however, for
computational expediency in these simulations we work in the limit of zero
thermal inertia, so that the absorbed radiation is instantaneously re-emitted.
Non-zero thermal inertia changes the obliquity torques, but not the spin
torques, so this simplification is a reasonable strategy for obtaining
statistically representative descriptions of spin evolution. The thermal
emission is assumed to be Lambertian (i.e., isotropic into the sky hemisphere),
with a correction for partial blockage of the sky by an elevated horizon
\citep{StatlerYORP}. The code computes the torques from both the reflected and
emitted radiation, though the latter dominates for typically dark asteroids.

\subsection{Self-Consistent Spin and Shape Evolution}

In order to self-consistently model the spin and shape evolution, we developed
four additional code elements that work with {\tt pdkgrav} and {\tt TACO}, and
carry out the following tasks:
\begin{enumerate}
\item
Fit a triangular tiling over a {\tt pkdgrav} object composed of spheres,
to pass to {\tt TACO} for computing the YORP torques;
\item
Identify when a movement of material has occured, and minimally adjust
the tiling to accommodate the movement (leaving it unchanged over the part of
the surface where no movement occurred);
\item
Integrate the spin and obliquity in time using the torques calculated by
TACO; and finally,
\item Orchestrate the entire procedure, running and passing data between the
codes.
\end{enumerate}
We describe each of these elements in detail below.

\subsubsection{Tiling}

\begin{figure}[t]
\centerline{ \epsfig{file=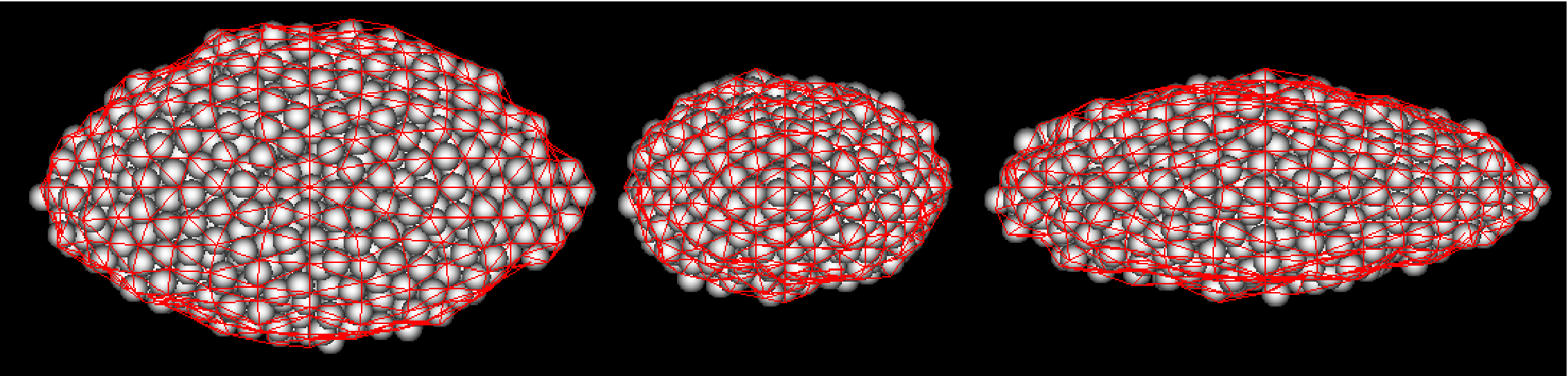, scale=0.42} }
\caption{Othogonal views of an aggregate of spheres along with the
triangular tiling in red, projected onto the ({\em left\/}) $xy$ plane,
({\em center\/}) $yz$ plane, ({\em right\/}) $xz$ plane.
\label{fig:tiling2}
}
\end{figure}

Our initial test objects (Section \ref{s.initialconditions}) are aggregates of
identical spheres. To fit a tiling over an object, we first compute and
diagonalize the intertia tensor, and rotate the object to principal axis
orientation with the center of mass at the origin and the $x$, $y$, and $z$
axes corresponding to the long, middle, and short axes, respectively.
We then create a tiling
of the equivalent ellipsoid with the same bulk density. At this point the
ellipsoidal tiling is close to the object, and the goal is to adjust the
vertices to fit the tiling tightly around it. We define the function
\begin{equation}
G(x,y,z) \equiv R^{2n} \sum_{j=1}^N
\left[(x-x_j)^2+(y-y_j)^2+(z-z_j)^2\right]^{-n}-G_0,
\end{equation}
where $R$ is the sphere radius, $(x_j,y_j,z_j)$ are the coordinates of the
center of sphere $j$, $N$ is the number of spheres, and $n$ and $G_0$ are constants chosen so that the
surface $G(x,y,z)=0$ tightly surrounds the object. We have found by trial and
error that the choice $n=2$ and $G_0=1.25$ works well for a variety of aggregate
shapes.  Each vertex of the ellipsoidal tiling is moved in or out in the
direction normal to the ellipsoid, to place it on the surface $G(x,y,z)=0$ as
shown in Fig.~\ref{fig:tiling2}.  Finally, the tiling is rotated
back to the orientation of the original object. 

One should remember that both the system of spheres and the triangular
tiling are numerical idealizations. They are intended to simulate the
collective behavior of a real aggregate composed of irregular rocks, pebbles,
and regolith, not to literally represent the constituent pieces. Hence
there is no need to resolve each sphere individually with an
extremely fine mesh, or to resolve each surface facet by filling the interior
with tiny spheres.

Nonetheless, \citet{StatlerYORP} emphasized the extreme sensitivity of YORP to
the detailed topography of asteroid surfaces. So we should be concerned about
the sensitivity of the computed torques in our simulations both to the
resolution of the tiling and to the positioning of the tiling on the aggregate
object. We have tested this by calculating the torques on a small selection of
aggregates at 9 different resolutions (determined by the number of tiles,
ranging from 784 to 19,960) and small angular shifts of the tilings (by a
few degrees). As expected, we find that the torque vaires typically by
tens of percent among the various shifts and resolutions. This result implies
that the exact results of our simulations will depend on arbitrary choices of
parameters related to the resolution and tiling. We adopt the lowest
resolution consistent with the number of spheres in the initial objects, and
stress that the detailed results of each simulation will be
resolution-dependent, and should be interpreted only as {\em examples\/} of
the types of behavior that may result from self-consistent YORP.

\subsubsection{Detecting Material Movement and Updating the Tiling}

\begin{figure}[t]
\centerline{ \epsfig{file=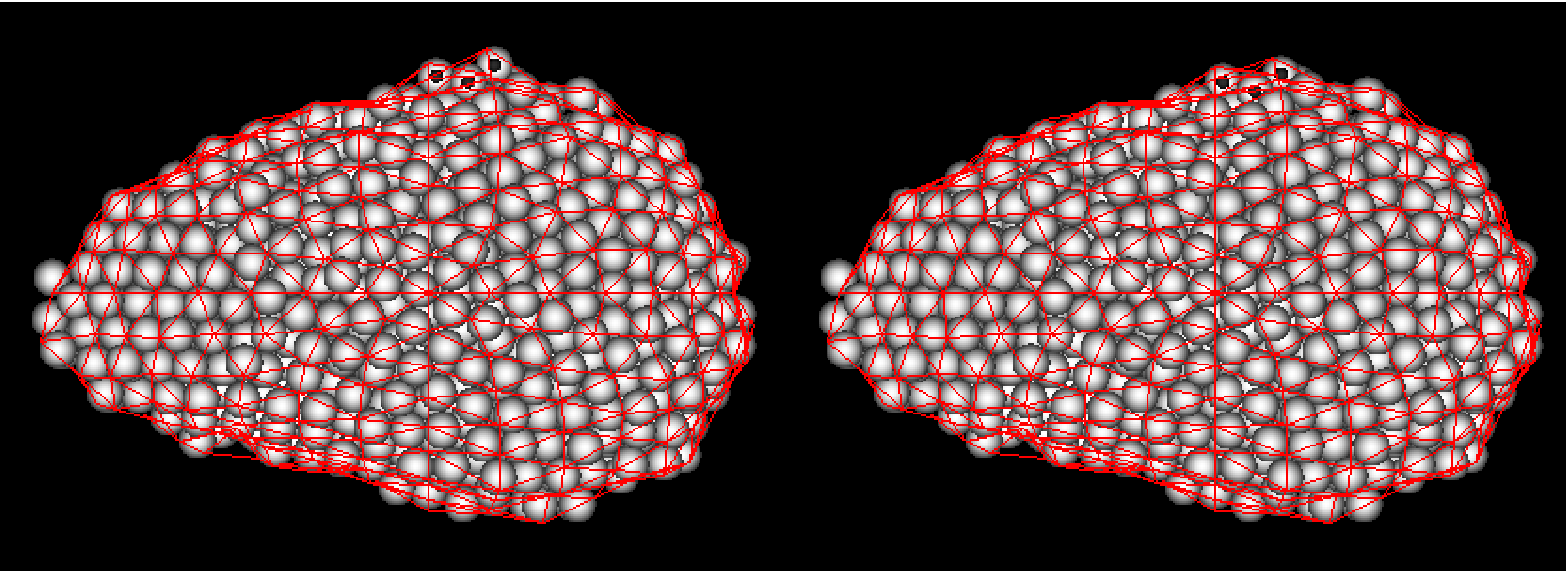, scale=0.6} }
\caption{Showing the adjustment of the tiling on an aggregate object before
({\em left\/}) and after ({\em right\/}) a small movement of
material. The line of sight is along the spin axis. The tiling is shown in red.
Black asterisks identify 3 spheres that move by more than a quarter of the
sphere's radius. Note that after the movement the tiling has been modified in
that area while remaining unchanged elsewhere.
\label{fig:tiling3}
}
\end{figure}

We define a movement of material as a shift of one or more {\tt pkdgrav}
spheres by more than a quarter of its radius. To determine whether
a movement has occurred, we compare the current object with the object
resulting from the previous movement. If no spheres have moved, the current
object should be a rotated and translated copy of the earlier object, except
for small differences caused by the slight bouncing of spheres that is
inherent in the HSDEM approach. We use the LMDIF routine from the MINPACK
\citep{min2} library to fit for the three Euler angles and three displacements
describing the rotation
and translation that minimizes the sum of the squares of the differences
in sphere positions. After the initial fit, the spheres that have moved
by more than the allowed tolerance are flagged and excluded, and the fit is
obtained again. The process is iterated until none of the remaining
spheres is flagged as having moved. Figure \ref{fig:tiling3}
shows an example of two consecutive objects, with the spheres identified
as having moved marked with black spots.

When a material movement has occurred, we need to update the tiling. However,
we must ensure that, insofar as is possible, the tiling is altered only
over the regions where motion occurred, so that any changes to the YORP
torques are due to the motion itself and are not merely the result of a shifted
tiling. To ``minimally evolve'' the tiling, we transform
the new object back to the {\em original\/} orientation at time $t=0$, and 
re-build the tiling starting from the {\em original\/} equivalent ellipsoid.
This guarantees that those spheres that do not move will be at the same
position that they were initially and therefore will receive the same tiling.
Figure \ref{fig:tiling3} shows the tilings on the example objects before
and after material motion.

Our minimal evolution algorithm can encounter difficulties when an
initially flattened or elongated object becomes significantly rounder.
As described below, this limits the simulations to 
objects with intitial flattenings $c/a > 0.5$.

\subsubsection{Spin-State Evolution}

The rate of change of the obliquity $\epsilon$ and the angular velocity
$\omega$ are given by \citep{Rub00}
\begin{equation}
\frac{d
\epsilon}{dt}=\frac{T_{\epsilon}}{C \omega} \label{eq1}
\end{equation}
and
\begin{equation}
\frac{d \omega}{dt}=\frac{T_{\omega}}{C} \label{eq2}
\end{equation}
where $C$ is the moment of inertia about the rotation axis and
$T_{\epsilon}$ and $T_{\omega}$ are spin- and orbit-averaged torque
components, respectively: $T_{\omega}$ is the component parallel
to the spin axis, and $T_{\epsilon}$ is the orthogonal component that lies in 
the plane containing both the spin axis and the orbit normal.
Once a tiling is obtained for a given object, 
$T_{\epsilon}$ and $T_{\omega}$ are calculated using TACO over an obliquity
grid with a spacing of $5\arcdeg$. At intermediate values of obliquity,
the torques are interpolated from the grid.
Equations (\ref{eq1}) and (\ref{eq2}) are
solved numerically using a fourth-order Runge Kutta integrator with a
$10^{3}$ year step size. We have
verified that this routine reproduces the exact analytic results for
idealized cases of rigid-body evolution in which the YORP curves
take the forms $T_{\omega} \propto \cos \epsilon$
and $T_{\epsilon} \propto \sin \epsilon$.

\subsubsection{Orchestrating the Simulations}

Top-level control of the simulations is handled by a {\tt python} script
that orchestrates the back-and-forth stepping between {\tt TACO} and
{\tt pkdgrav} and enables their interaction with the
additional routines described above. Details of the logic, including a
flowchart showing the individual steps, are given in the Appendix.

\subsection{Initial Conditions}\label{s.initialconditions}

We select our initial test objects from a collection of 144 rotating
equilibria created using {\tt pkdgrav} \citep{Tan09}. These authors built
ellipsoidal aggregates with various shapes and initial spins, and then
allowed them to evolve and reconfigure dynamically until they reached stable
configurations. The objects have a natural disordered packing, and are
composed of 1000 spheres of radius $50.2\m$, each with density of
$2.96\g\cm^{-3}$. The bulk densities and mean diameters are in the range of
$1.55$ to $1.72\g\cm^{-3}$ and $1.3$ to $2.0\km$, respectively. We tile each
object, compute
the torques, and integrate the spin state evolution it would undergo if
it remained rigid. We intentionally pick objects that, were they
rigid bodies, would initially accelerate in spin rate and
 display a representative range of YORP-cycle behaviors.
In particular, we select four objects that would spin up at all obliquities,
with rigid-body end states in which (formally) $\omega \to \infty$ as
$t \to \infty$. The remaining 12 objects are chosen to be approximately
uniformly distributed in the axis-ratio plane, subject to the requirement
that $c/a \ge 0.5$ to avoid numerical difficulties in minimally evolving the
tiling.

\begin{figure}[t]
\centerline {\includegraphics[scale=0.50]{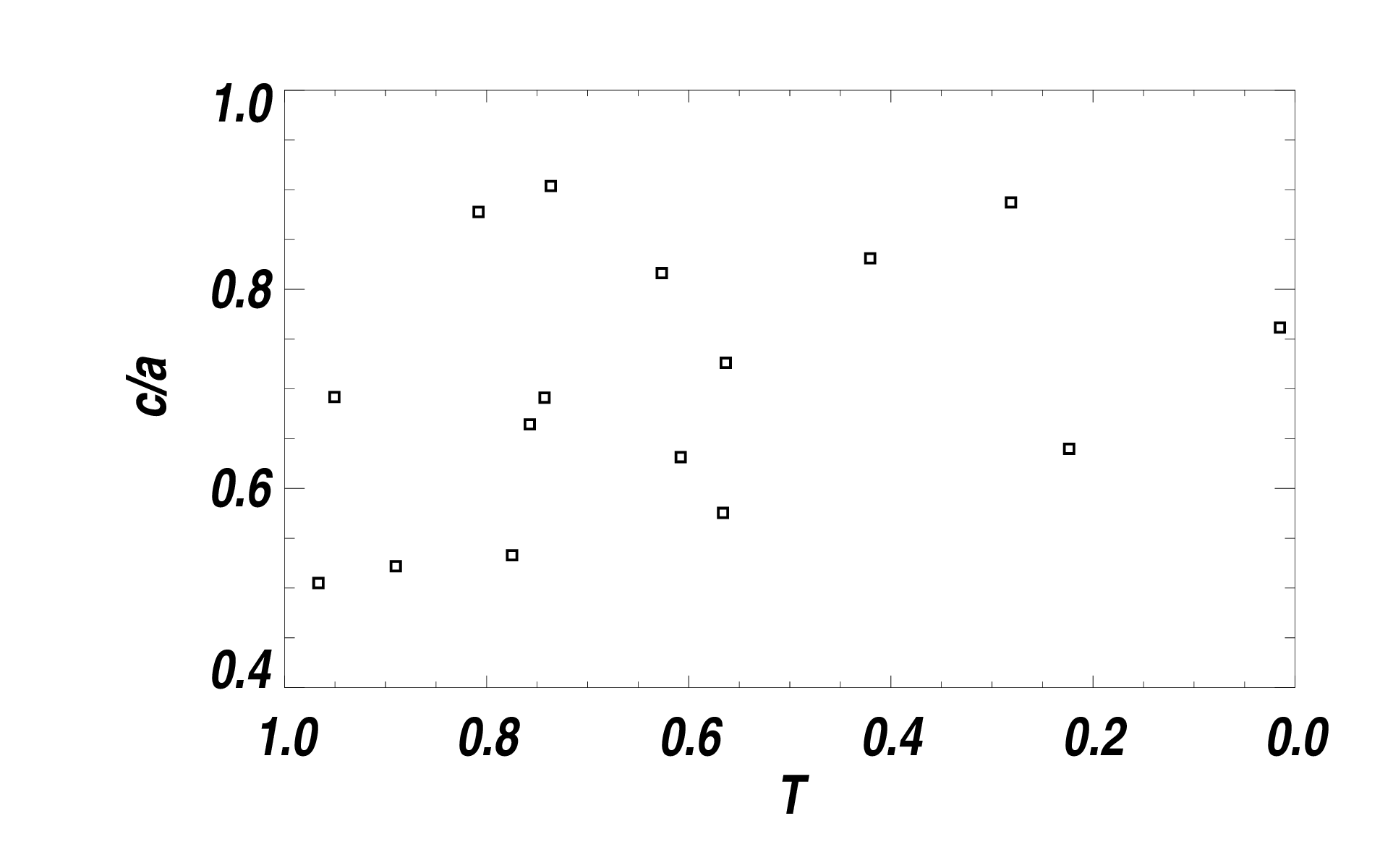}}
\caption{
The semi-axis ratio distribution of the initial shapes, plotted in
terms of the short axis ratio $c/a$ and
the triaxiality parameter $T \equiv [1-(b/a)^2]/[1-(c/a)^2]$.
Purely oblate objects have $T=0$ and purely prolate objects have $T=1$.
\label{fig:initialdisorig}
}
\end{figure}

Table \ref{tab:aggregates} shows the initial parameters for our sample of 16
aggregates. Figure \ref{fig:initialdisorig} shows the
initial distribution of shapes in the axis-ratio plane, plotted in terms of
the short-to-long axis ratio ($c/a$) and the
triaxiality parameter $T$, defined\footnote{This definition of $T$ matches 
that used in galaxy dynamics \citep[e.g.,][]{StatlerNGC4365}.}\ by
\beq
T \equiv {{1 - (b/a)^2} \over {1 - (c/a)^2}}.
\label{eq:triax}
\eeq
Objects with $T=0$ are oblate spheroids ($b=a$), and those with $T=1$ are prolate
spheroids ($b=c$). The initial distribution
is representative of the distribution of known asteroid shapes
approximated by triaxial ellipsoids \citep{Tan09, Kry07}.

\begin{table}[t]
\centering
\caption{Initial Aggregate Objects}
\label{tab:aggregates}
\begin{tabular}{*6c}
\hline
Simulation &  \multicolumn{2}{c}{Semi-axis ratio} & Semi-major axis $a$ &
Bulk Density & Period \\
{}   & {$b/a$} & {$c/a$}  & (km)    & ($\g\cm^{-3}$)   & (hours) \\
\hline
 1 & 0.91  &  0.88  &  0.686  &  1.55  &  10.08 \\
 2 & 0.88  &  0.87  &  0.696  &  1.66  &  10.00 \\
 3 & 0.94  &  0.83  &  0.688  &  1.62  &  10.51 \\
 4 & 0.86  &  0.74  &  0.722  &  1.62  &\po5.52 \\
 5 & 0.88  &  0.82  &  0.701  &  1.66  &  10.34 \\
 6 & 0.78  &  0.69  &  0.765  &  1.66  &\po5.72 \\
 7 & 0.70  &  0.70  &  0.797  &  1.66  &\po5.74 \\
 8 & 0.74  &  0.65  &  0.799  &  1.63  &\po6.09 \\
 9 & 0.77  &  0.62  &  0.786  &  1.67  &\po6.26 \\
10 & 0.99  &  0.76  &  0.698  &  1.61  &\po5.70 \\
11 & 0.97  &  0.89  &  0.672  &  1.72  &\po5.00 \\
12 & 0.51  &  0.50  &  1.000  &  1.60  &\po5.28 \\
13 & 0.68  &  0.55  &  0.871  &  1.56  &\po4.54 \\
14 & 0.59  &  0.53  &  0.935  &  1.55  &\po4.84 \\
15 & 0.78  &  0.59  &  0.795  &  1.68  &\po4.35 \\
16 & 0.92  &  0.64  &  0.745  &  1.63  &\po4.19 \\
\hline
\end{tabular} 
\end{table}

We dynamically evolve the initial objects with {\tt pkdgrav} for several
rotations to ensure that all mass motion has stopped. We then recompute the
initial tilings and torques, and again consider the rigid body evolution of
the fully settled objects. We choose initial obliquities so that, if they
remained rigid, the objects would evolve through a wide range of obliquities
and spin rates. The adopted initial obliquity, the spin rate and obliquity
at the end of the YORP cycle (which we refer to as the rigid-body end state),
and the time $t_{\rm YC}$ required to complete the cycle are listed
for each object in Tables \ref{tab:spinrates} and \ref{tab:obliquity}.
Owing to shape adjustments during the initial settling,
only 2 of the 16 objects have rigid-body end states involving continual,
indefinite spin-up, and the object in simulation 16 is initially decelerating.

\subsection{Choice of Code Parameters}

\begin{itemize}
\item
The normal and tangential coefficients of restitution in {\tt pkdgrav},
$\varepsilon_n$ and $\varepsilon_t$, are set to 0.2 and 0.5 for all
simulations.  These values were chosen in order to ensure a fair amount of
dissipation given the compression of the timescales for the forces considered
here.  Larger values of $\varepsilon_n$ and $\varepsilon_t$ would result in
the need of longer timescales to damp the particle motions, but in practice
most particle motions are so small that the precise choice of these
parameters makes little difference.
Similarly, our choices of slide limit ($0.01$ times the particle mutual
escape speed) and inelastic collapse limit ($1\times 10^{-5}$ in dimensionless
units) are relatively conservative to encourage dissipation but still
avoid numerical problems for small particle motions with HSDEM.  Since we
expect YORP timescales typically to be longer than dissipative timescales,
we do not expect that these choices will greatly affect our major results. 
However, we will not be able to constrain the real setttling times or how
long the objects might stay in non-principal axis spin state after a mass
movement.

\item
We adopt the lowest resolution in {\tt TACO} (784 tiles) consistent with the
number of spheres (1000) in the initial objects. Despite the sinsitivity
of the YORP torques to the details of the tiling, we expect
statistical results, such as the fraction of objects exhibiting various types
of behavior, to be relatively robust. We
re-run a subset of the simulations at twice the linear resolution (3184
tiles) to verify this expectation.
\item
We adopt the Hapke model parameters for an average S-type asteroid determined
by \citet{Hel89}: a single-scattering albedo of $w=0.23$, a surface roughness
or mean slope angle of $\bar{\theta}=20^\circ$, and an asymmetry parameter
$\xi=-0.35$. The opposition effect is neglected.
\item
As explained above, we set the thermal inertia to zero, so that the
absorbed radiation is re-emitted instantaneously. Since a non-zero thermal
inertia alters the obliquity torques, and not the spin torques, we adopt
this strategy for obtaining statistically representative
results for how the spin state evolution of aggregates compares to that of
rigid bodies {\em under the same assumptions.} Neither the rigid-body nor
the aggregate evolution simulated here will reproduce the known tendency for
YORP to drive objects toward obliquities of $0\arcdeg$ and $180\arcdeg$, which
is largely a consequence of finite thermal inertia \citep{Cap04}.
\item
We assume all objects are in circular orbits around the Sun at 1 AU.
\end{itemize}

Nearly all simulations are run initially to a time of $15\myr$, as the typical
dynamical lifetimes of NEAs are around $10\myr$. Simulations are continued
further if the rigid-body YORP cycle time $t_{\rm YC} > 15 \myr$. Some
simulations are terminated early if objects are spinning down toward zero
with slow rotation periods of over 20 hours. Objects for which rigid-body
YORP predicts infinite spin up in infinite time are run for $30\myr$.

\begin{table}[p]
{\centering
\caption{Summary of Spin Rate Evolution}
\label{tab:spinrates}
\begin{tabular}{ccrccrcrlrc}
\hline
{} & {} & \multicolumn{3}{c}{---\ Rigid-Body\ ---} & {} &
\multicolumn{3}{c}{---\ Aggregate\ ---}\\
Simulation & Initial$^{\rm a}$ & Max$^{\rm a}$ &  Min$^{\rm a}$  &
End$^{\rm a}$ & $t_{\rm YC}$$^{\rm b}$&
Max$^{\rm a}$ &  Min$^{\rm a}$ & End$^{\rm a,c}$ & $t_{\rm sim}$$^{\rm d}$
& Ev. Type$^{\rm e}$\\
\hline
\multicolumn{3}{l}{\it Standard resolution tiling:}\\
 1 & 2.4 &  9.1 & 0.0 & 0.0  &  3.6 & 4.7 & 0.8  & 0.8           &  3.2 & 
	MYC/Stg\\
 2 & 2.4 &  8.3 & 0.0 & 0.0  &  4.2 & 4.8 & 0.3  & 0.3$\searrow$ &  1.9 & 
	MYC\\
 3 & 2.3 &  7.3 & 0.0 & 0.0  &  1.7 & 5.6 & 0.2  & 0.2$\searrow$ &  3.5 & 
	MYC\\
 4 & 4.3 & 13.7 & 0.0 & 0.0  &  6.8 & 6.2 & 4.1  & 4.1$\leadsto$ & 15.0 & 
	Sto\\
 5 & 2.3 &  8.3 & 0.0 & 0.0  &  3.7 & 6.2 & 0.1  & 0.1$\searrow$ &  1.7 & 
	MYC\\
 6 & 4.2 & 26.8 & 0.0 & 0.0  &  7.3 & 6.0 & 4.2  & 4.7$\leadsto$ & 20.4 & 
	Sto/SG\\
 7 & 4.2 &  6.8 & 0.0 & 0.0  &  1.1 & 5.5 & 0.6  & 0.6           &  2.5 & 
	Sto/Stg\\
 8 & 3.9 & 22.6 & 0.0 & 0.0  & 16.0 & 6.1 & 0.4  & 3.9$\leadsto$ &  2.5 & 
	Sto/F\\
 9 & 3.8 & 25.8 & 0.0 & 0.0  &  9.4 & 6.1 & 0.5  & 0.5$\searrow$ &  3.3 & 
	Sto\\
10 & 4.2 & 20.1 & 0.0 & 0.0  &  5.3 & 6.4 & 4.2  & 5.1$\leadsto$ & 15.0 & 
	Sto\\
11 & 4.8 & 17.7 & 0.0 & 0.0  &  6.9 & 6.4 & 4.3  & 5.3$\leadsto$ & 30.0 & 
	Sto/Stg\\
12 & 4.6 & 12.2 & 0.0 & 0.0  &  3.2 & 5.4 & 3.8  & 4.1$\leadsto$ & 15.1 & 
	Sto/SG\\
13 & 5.3 & 11.7 & 0.0 & 0.0  &  2.5 & 5.9 & 0.3  & 0.5$\searrow$ &  7.0 & 
	Sto\\
14 & 4.9 & Inf. & 4.9 & Inf. & Inf. & 5.6 & 4.8  & 5.4$\leadsto$ & 30.0 & 
	Sto\\
15 & 5.5 & Inf. & 5.5 & Inf. & Inf. & 6.1 & 4.4  & 4.5$\leadsto$ & 31.0 & 
	Sto/SG/Stg\\
16 & 5.7 &  5.7 & 0.0 & 0.0  &  2.9 & 5.7 & 0.0  & 0.0           &  1.2 & 
	MYC\\
\multicolumn{3}{l}{\it High resolution tiling:}\\
 6H & 4.2 & Inf. & 4.2 & Inf. & Inf. & 6.0 & 0.7 & 0.7$\searrow$ & 15.8 & Sto\\
 8H & 3.9 & 6.5  & 0.0 & 0.0  &  4.8 & 6.1 & 3.4 & 5.1$\leadsto$ & 13.5 & Sto\\
10H & 4.2 & 25.5 & 0.0 & 0.0  & 12.1 & 6.6 & 0.8 & 0.8$\leadsto$ & 13.2 & Sto\\
13H & 5.3 & Inf. & 5.2 & Inf. & Inf. & 5.9 & 3.9 & 4.6$\leadsto$ & 30.0 & Sto\\
\hline\\
\end{tabular}\par}
$^{\rm a}$ Spin rates in revolutions day $^{-1}$.\\
$^{\rm b}$ YORP cycle completion time in Myr.\\
$^{\rm c}$ Symbol indicates trend at simulation end:
	$\nearrow$ increasing; $\searrow$ decreasing; $\leadsto$ varying;
	no symbol: constant\\
$^{\rm d}$ Simulation duration in Myr.\\
$^{\rm e}$ Descriptive classification of spin evolution:
``MYC'' = modified YORP cycle;
``Sto'' = stochastic;
``SG'' = self-governed;
``Stg'' = stagnating;
``F'' = ending with fission event.
\end{table}

\begin{table}[p]
{\centering
\caption{Summary of Obliquity Evolution}
\label{tab:obliquity}
\begin{tabular}{ccclc}
\hline
{} & {} & Rigid-Body & \multicolumn{2}{l}{\ \ \ ---\ Aggregate\ ---}\\
Simulation & Initial$^{\rm a}$ & 
End$^{\rm a}$ &
End$^{\rm a,b}$ &
Ev. Type$^{\rm c}$\\
\hline
\multicolumn{3}{l}{\it Standard resolution tiling:}\\
 1 &  5 & 90 & 90           & MYC\\
 2 &  5 & 90 & 90           & MYC\\
 3 &  5 & 90 & 83$\nearrow$ & Sto\\
 4 &  5 & 90 & 50$\leadsto$ & Sto\\
 5 &  5 & 90 & 90           & MYC\\
 6 &  5 & 90 & \phantom{9}0 & Sto\\
 7 &  5 & 90 & 80           & MYC\\
 8 &  5 & 90 & 19$\nearrow$ & Sto/F\\
 9 &  5 & 90 & 90           & Sto\\
10 &  5 & 90 & \phantom{9}0 & Sto\\
11 &  5 & 90 & 12$\leadsto$ & Sto/Stg\\
12 &  5 & 90 & 23$\leadsto$ & Sto/SG\\
13 &  5 & 90 & 90           & Sto\\
14 &  5 & 90 & \phantom{9}6$\leadsto$ & Sto\\
15 &  5 & 90 & 30$\leadsto$ & Sto/SG/Stg\\
16 & 85~ & 86 & 90           & MYC\\
\multicolumn{3}{l}{\it High resolution tiling:}\\
 6H & 5 & 73 & 66$\leadsto$ & Sto\\
 8H & 5 & 86 & \phantom{9}2$\nearrow$ & Sto/Stg\\
10H & 5 & 85 & \phantom{9}2$\nearrow$ & Sto/Stg\\
13H & 5 & 85 &  20.6$\searrow$            &Sto/Stg\\
\hline\\
\end{tabular}\par}
$^{\rm a}$ Obliquities in degrees.\\
$^{\rm b}$ Symbol indicates trend at simulation end:
	$\nearrow$ increasing; $\searrow$ decreasing; $\leadsto$ varying;
	no symbol: constant\\
$^{\rm c}$ Descriptive classification of spin evolution:
``MYC'' = modified YORP cycle;
``Sto'' = stochastic;
``SG'' = self-governed;
``Stg'' = stagnating;
``F'' = ending with fission event.
\end{table}

\section{Results}\label{s.results}

\subsection{YORP Self-Limitation}

\begin{figure}[t]
\centerline{ \includegraphics[scale=0.55]{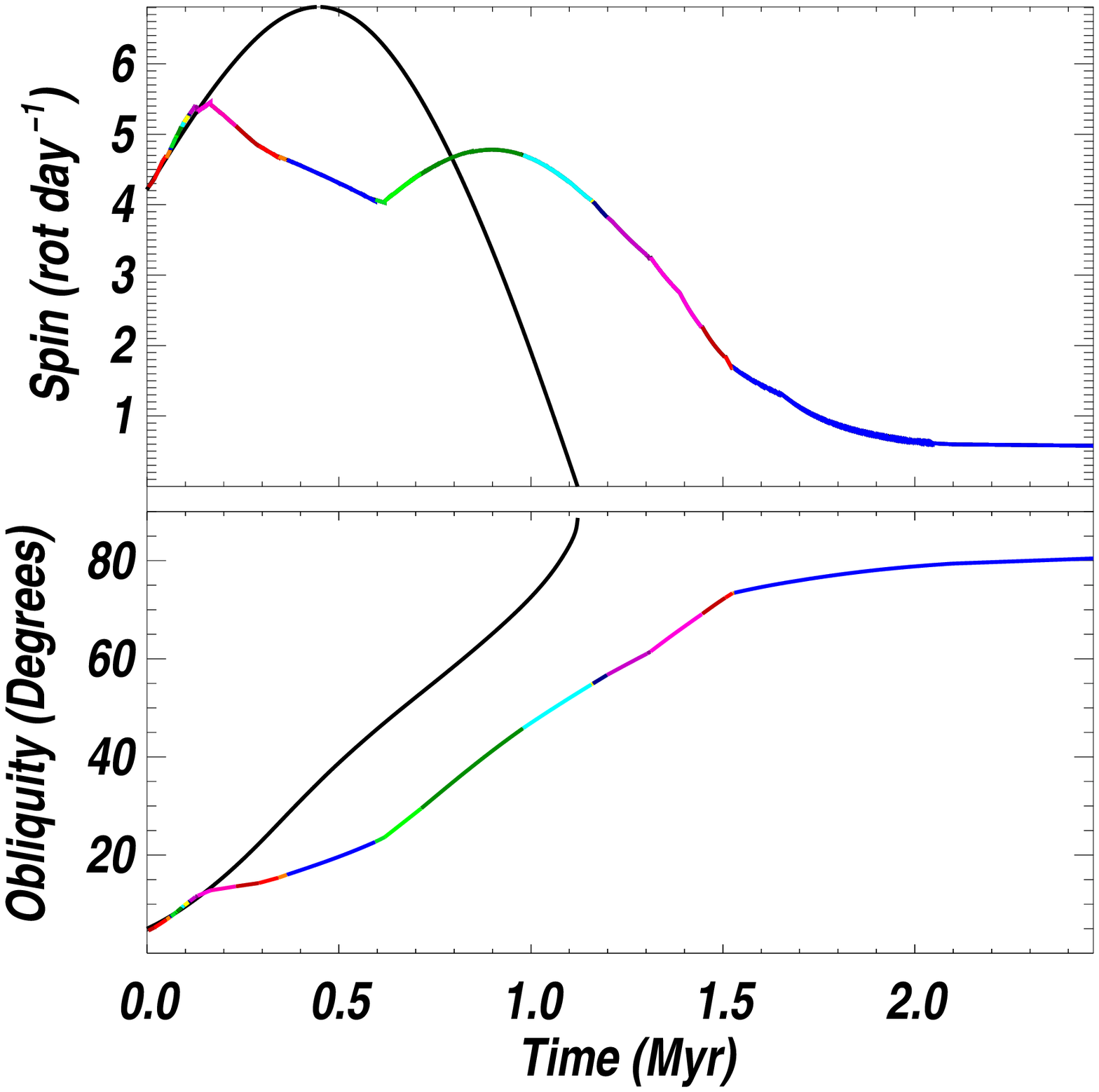} }
\caption{
Spin ({\em top\/}) and obliquity ({\em bottom\/}) evolutions in simulation
7. {\em Black lines\/} show the rigid-body evolution. {\em Colored lines\/}
show the actual evolution of the aggregate; each color represents a new shape
with a corresponding torque. The spin evolution is weakly stochastic, resulting
in mild YORP self-limitation. The spin reaches 80\% of the maximum rigid-body
rate and eventually stagnates at a $\sim 14\hour$ period. The obliquity
changes monotonically and more resembles a YORP cycle, but asymptotes to an
unusual end state of $80\arcdeg$.
\label{fig:newfig8_sim10}
}
\end{figure}

\begin{figure}[t]
\centerline{ \includegraphics[scale=0.55]{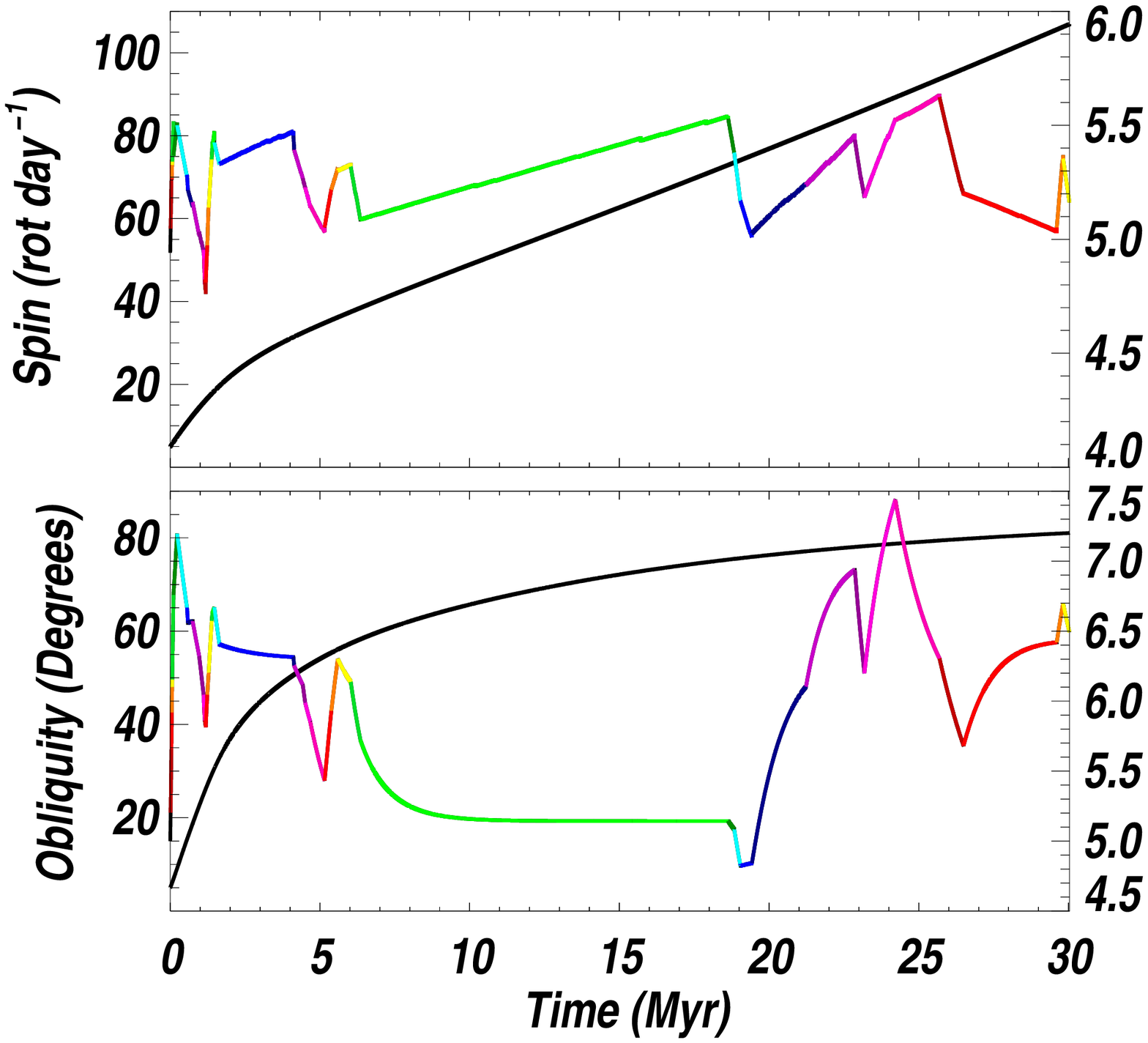} }
\caption{
Spin ({\em top\/}) and obliquity ({\em bottom\/}) evolutions in simulation
14, similar to Fig.~\protect{\ref{fig:newfig8_sim10}}.
{\em Black lines\/} and {\em left scales\/} show the rigid-body evolution.
{\em Colored lines\/} and {\em right scales\/} show the evolution of the
aggregate. Both spin and obliquity evolutions are highly stochastic and
result in strong self-limiation.
\label{fig:newfig6_sim20}
}
\end{figure}

\begin{figure}[t]
\centerline{ \includegraphics[scale=0.55]{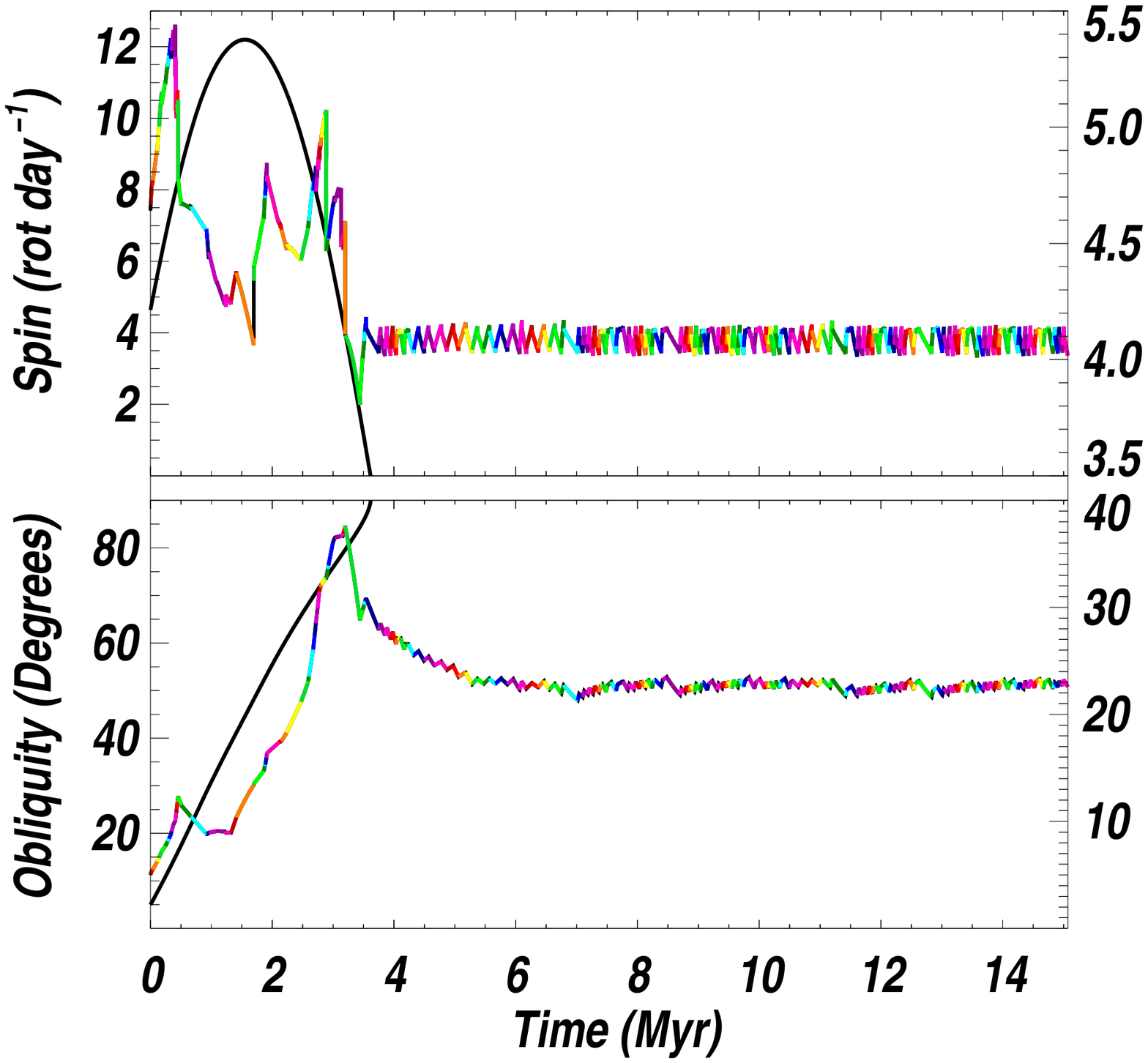} }
\caption{
Spin ({\em top\/}) and obliquity ({\em bottom\/}) evolutions in simulation 12.
{\em Black lines\/} and {\em left scales\/} show the rigid-body evolution.
{\em Colored lines\/} and {\em right scales\/} show the evolution of the
aggregate.  The evolution is stochastic for the first $3.7\myr$, after which
it becomes self-governed, toggling between neighboring configurations that
alternately accelerate and decelerate the spin.
\label{fig:newfig9_sim18}
}
\end{figure}

\begin{figure}[t]
\centerline{ \includegraphics[scale=0.55]{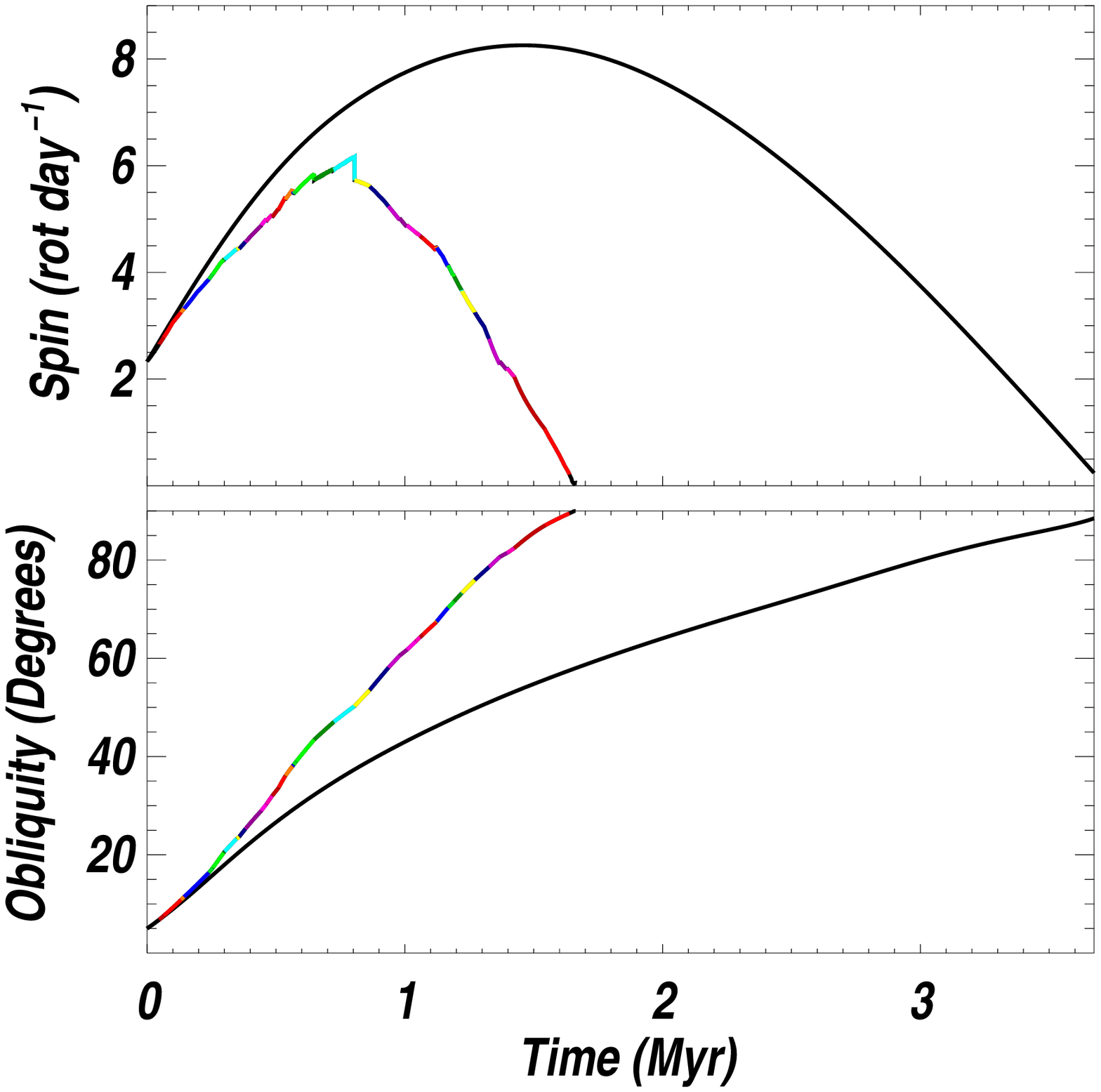} }
\caption{
Spin ({\em top\/}) and obliquity ({\em bottom\/}) evolutions in simulation 5,
as in Fig.~\protect{\ref{fig:newfig8_sim10}}. Only one change in the sign
of the spin component of torque occurs, resulting in evolution that
resembles a modified YORP cycle.
\label{fig:newfig_sim7}
}
\end{figure}

The time evolution of the rotation rate and obliquity in a representative
selection of our simulations is shown in Figures \ref{fig:newfig8_sim10},
\ref{fig:newfig6_sim20}, \ref{fig:newfig9_sim18}, and \ref{fig:newfig_sim7}.
In each figure, the solid black lines show the evolution expected if the
object had remained rigid, while the actual evolution of the aggregate
is shown in a color sequence. Each color corresponds to a new configuration,
and every change of color corresponds to a movement of material requiring a
recalculation of the torques. Every object simulated undergoes multiple changes
in shape, and no aggregate evolves according to the rigid-body prediction.

Tables \ref{tab:spinrates} and \ref{tab:obliquity} summarize the evolutions
in spin rate and obliquity, respectively. The most robust and striking result
is the narrow range of spin rates attained by the evolving aggregates
compared with their rigid counterparts. Column 3 in the upper section of Table
\ref{tab:spinrates} shows that the ordinary YORP cycle would have accelerated
9 of the 16 rigid bodies (at the standard {\tt TACO} resolution) past the
nominal $2\hour$ rubble-pile spin limit, and 4 of them to periods shorter
than $1\hour$. As rigid bodies, every object but one would have reached
maximum spin rates faster than $6.5\,\mbox{\rm rot}\,\mbox{\rm day}^{-1}$. But
as aggregates, not a single one ever spins this fast. As rigid bodies,
all objects but two would subsequently have spun down to zero in times ranging
from $1.1$ to $16\myr$. As aggregates, only 5 objects spin down effectively
to zero, or are headed that way at the end of the simulation. Of the remainder,
7 are still spinning at rates $> 4\,\mbox{\rm rot}\,\mbox{\rm day}^{-1}$,
3 are rotating slowly at $< 1\,\mbox{\rm rot}\,\mbox{\rm day}^{-1}$,
and one has fissioned (about which more below). The lower part of Table
\ref{tab:spinrates} confirms that these same qualitative results regarding
maximum and minimum spin rates hold
in the simulations rerun at higher {\tt TACO} resolution.

Aggregate bodies thus resist---and avoid---the wide excursions in spin rate
implied by the rigid-body YORP cycle. Because the resistance is produced by the
YORP-driven deformation of the object, we refer to this overall
phenomenon as {\em YORP self-limitation\/}, or {\em self-limited YORP.}

We observe three distinct behaviors that can give rise to YORP self-limitation:
\begin{itemize}
\item
{\em Stochastic YORP\/}, in which the object random-walks among different
shape configurations, resulting in a sequence of episodes of unpredictable
duration, each resembling part of a YORP cycle;
\item
{\em Self-Governing YORP\/}, in which the object toggles between a small number
of configurations, resulting in a limit cycle that restricts the
spin and obliquity to a narrow range; and
\item
{\em Stagnating YORP\/}, in which the object settles into a long-lived
configuration of very low torque well before reaching a YORP cycle end-state.
\end{itemize}
An object can exhibit any of these behaviors in its spin or obliquity
evolution. Spin and obliquity do
not need to behave in the same way; and multiple behaviors at different times
for a single object are common. 

The objects that do not exhibit YORP self-limitation (in either spin or
obliquity) as a result of one of the above behaviors are best described
as following a:
\begin{itemize}
\item
{\em Modified YORP Cycle\/}, which qualitatively resembles the typical YORP
cycle prediction, and in which changes in shape do not alter the direction
of evolution.
\end{itemize}
The last columns of Table \ref{tab:spinrates}
and Table \ref{tab:obliquity} indicate
the behaviors in spin and obliquity seen in each of the simulations.
We describe each of these four behaviors in more detail in the paragraphs below.
 
\begin{figure}[t]
\centerline{ \includegraphics[scale=0.55]{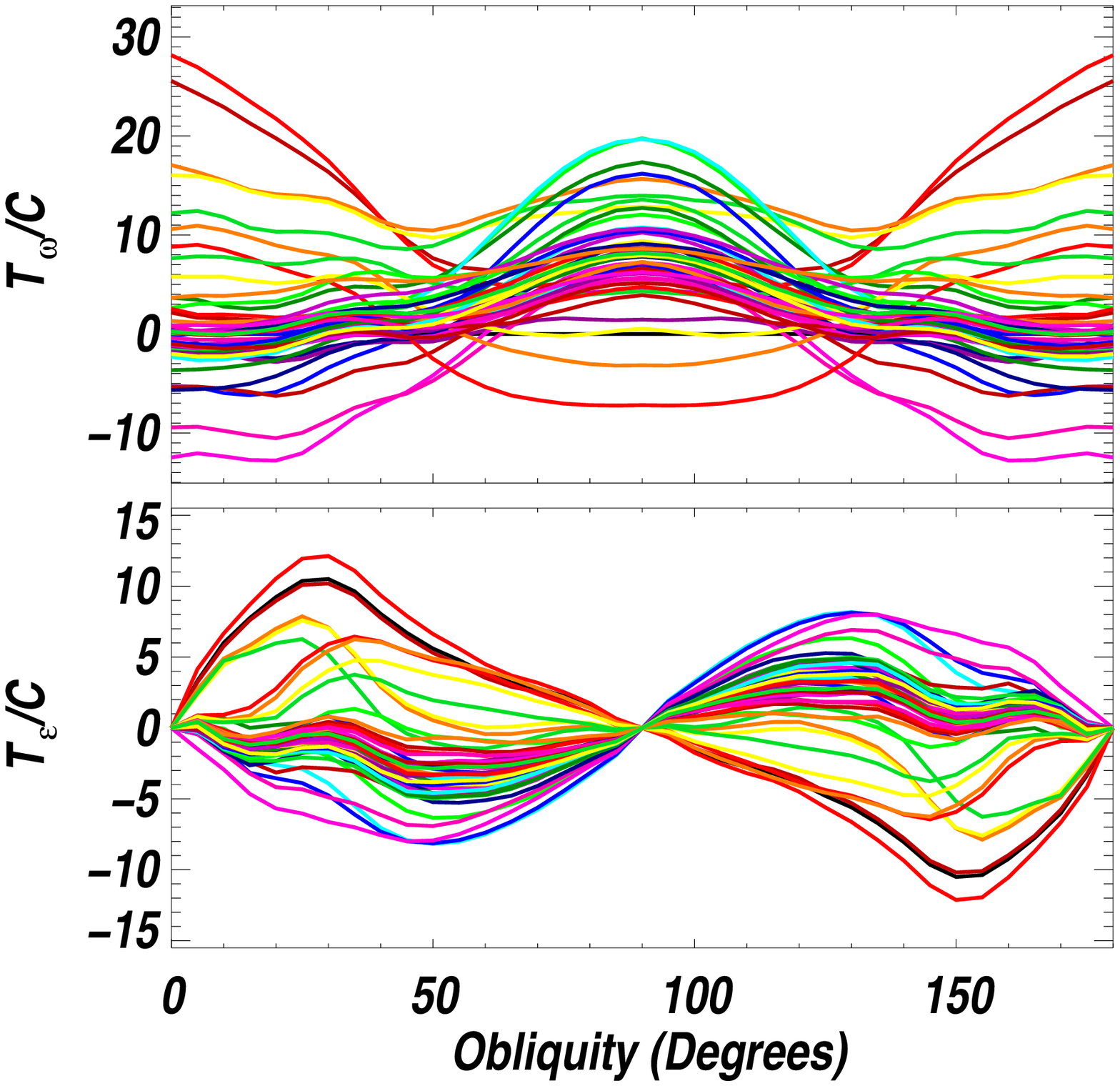} }
\caption{
The obliquity (bottom) and the spin (top) torques (YORP curves)
through which the aggregate
object in simulation 14 (Fig.~\protect{\ref{fig:newfig6_sim20}}) evolves.
The torque components divided by the moment of inertia (in units of
$10^{-18}\s^{-2}$) are plotted against the obliquity in degrees. Each colored
curve corresponds to a new shape of the aggregate object after a movement of
material, and corresponds to the segment of the same color in
Fig.~\protect{\ref{fig:newfig6_sim20}}.
\label{fig:newfig_torques}
}
\end{figure}

\subsubsection{Stochastic YORP}

Eleven of the 16 objects exhibit stochastic YORP in their spin evolution,
and an equal number (though not exactly the same objects) do so
in their obliquities.
The upper panel of Fig.~\ref{fig:newfig8_sim10} shows an example of weak
stochasticity in the spin evolution of the object in simulation 7. The
evolution has qualitative similarities to the YORP cycle prediction shown in
black, and many of the movements of material have only a slight effect on
the YORP torques. Nonetheless, the spin evolution changes direction multiple
times due to changes in the shape of the object. The obliquity evolution,
shown in the lower panel, is monotonic, with greater similarity to a YORP
cycle (which we discuss in section \ref{s.modyc} below), demonstrating that
different
types of YORP behavior can be seen in a single object at the same time.

An example of strongly stochastic YORP is shown in
Fig.~\ref{fig:newfig6_sim20}. Here, nearly every change in shape results in a
significant change in both components of torque, and often a change in their
signs. The scale of these changes can best be seen by looking at the sequence
of YORP curves that describe the shapes through which the object evolves.
This sequence is shown in Fig.~\ref{fig:newfig_torques}; keep in mind that
the object evolves along only a small fraction of each pair of YORP curves
before shifting to a new pair. As a result of these shifts, strong YORP
self-limitation confines spin and obliquity
to narrow intervals. Note that one can discern a few longer-lived
YORP-cycle-like episodes in Fig.~\ref{fig:newfig6_sim20}
(e.g., between 7 and $18\myr$); but the time
variability is non-repeating and unpredictable.

One can think of stochastic YORP as arising from two coupled effects. First
is the shape evolution itself, which causes the object to random-walk among
topographic configurations, each producing different YORP torques. Second is
the natural tendency for an evolving object to spend more time in
configurations that produce smaller torques, simply because it takes
longer to build up a sufficient change in spin to trigger a reconfiguration. 
Hence some points in the topographic space are ``stickier'' than others, and
the time an object may dwell in each configuration is a function
of the nearby topographic landscape and its past history. ``Sticky'' low-torque
configurations are also the cause of YORP stagnation, which we discuss below.

\begin{figure}[t]
\centerline{ \includegraphics[scale=0.35]{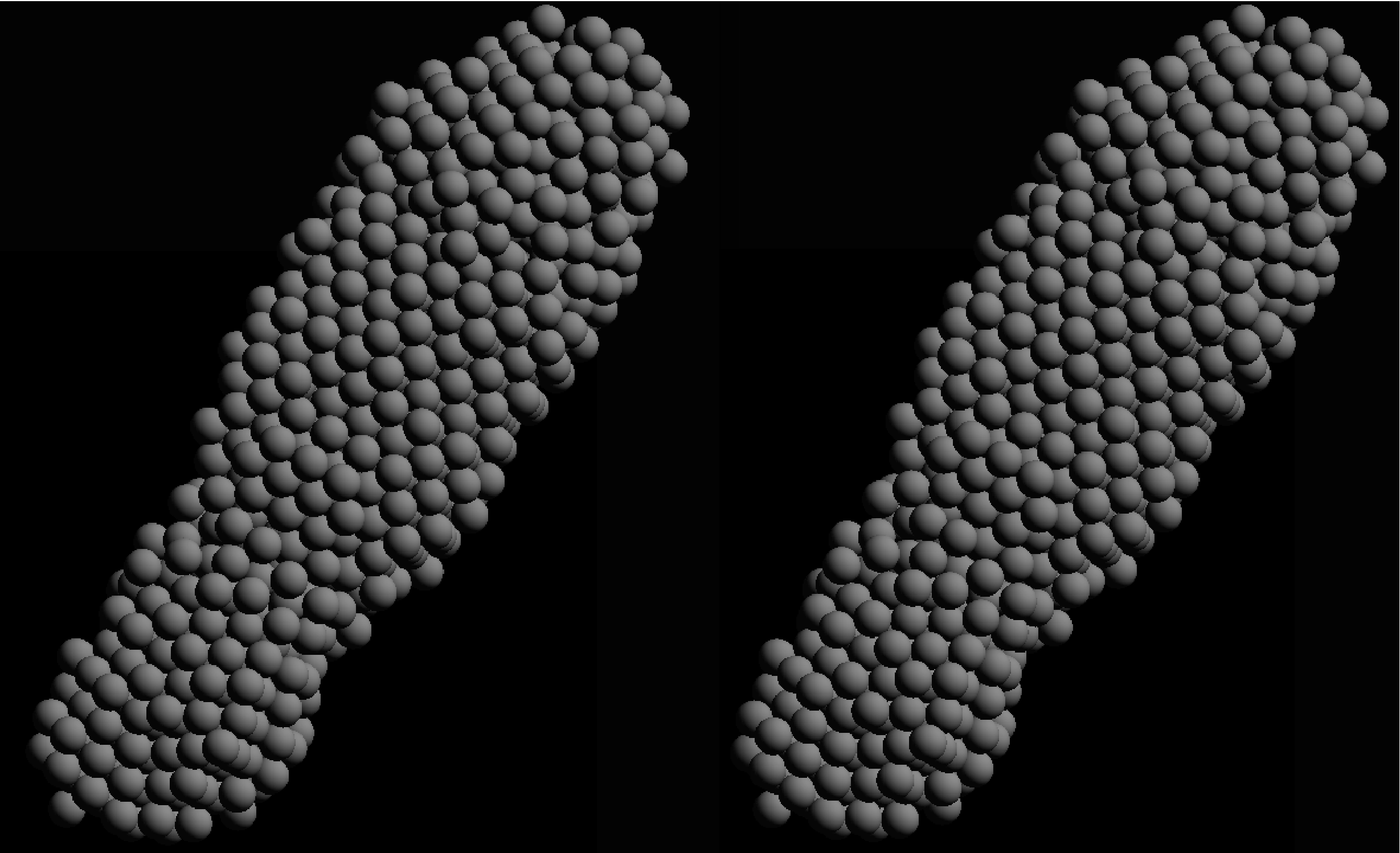} }
\caption{
The two configurations between which the object in simulation 12 toggles in
the self-governed phase of its evolution
(Fig.~\protect{\ref{fig:newfig9_sim18}}). The view is along the rotation
axis. Notice that the object bends at the constriction, one third of the
way up from the bottom.
\label{fig:sim12selfgovshapes}
}
\end{figure}

\begin{figure}[t]
\centerline{ \includegraphics[scale=0.55]{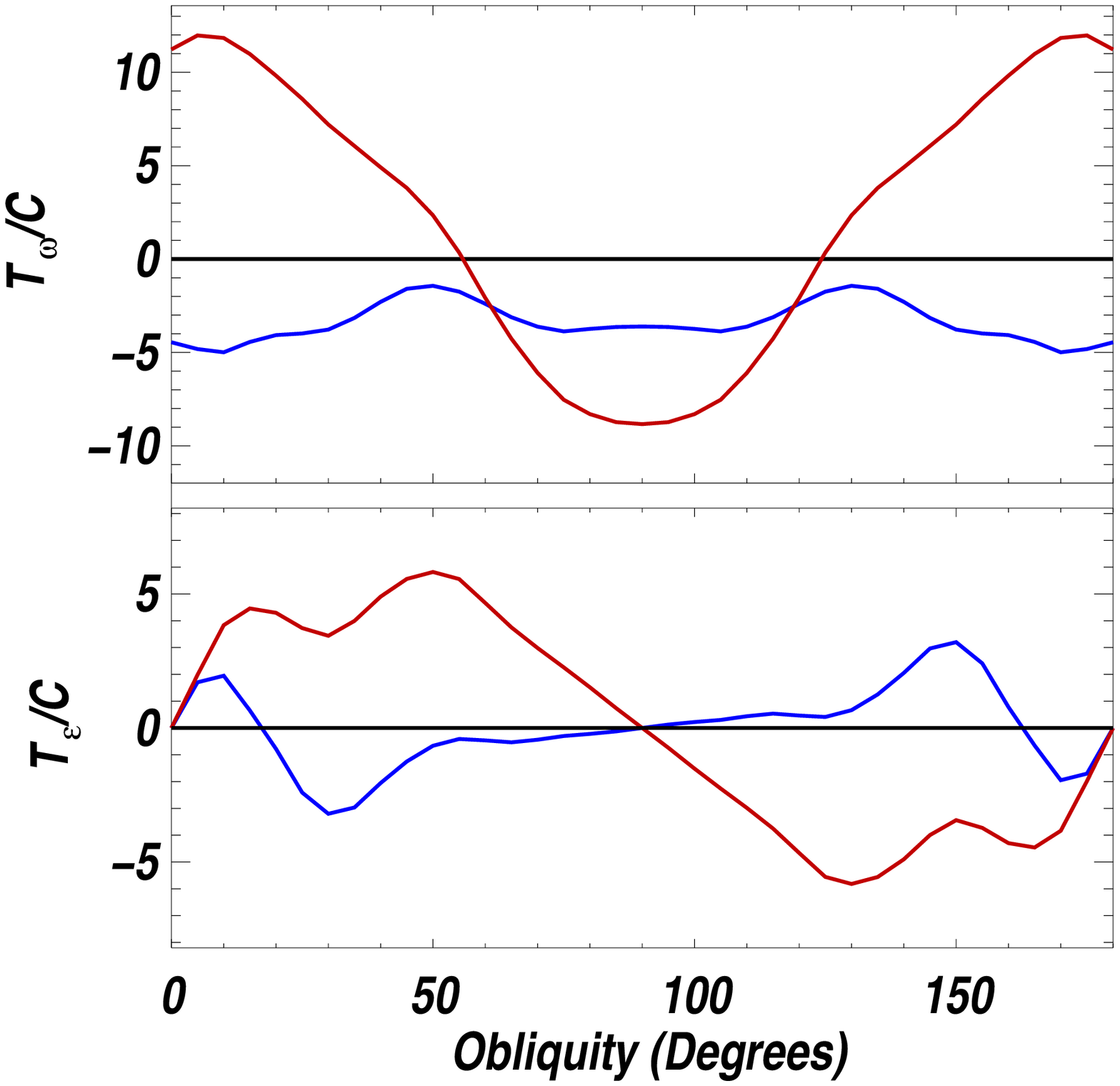} }
\caption{
Components of torque (YORP curves) generating changes in ({\em top\/}) spin and
({\em bottom\/}) obliquity for the two toggling configurations in simulation
12 (Fig.~\protect{\ref{fig:sim12selfgovshapes}}).
\label{fig:sim12selfgovtorques}
}
\end{figure}

\subsubsection{Self-Governed YORP}

Three objects show self-governing behavior in their spins, and two of these
are also self-governed in obliquity (the third having already evolved to an
$\epsilon = 0$ orientation before self-governing begins). Figure
\ref{fig:newfig9_sim18} shows simulation 12, which evolves stochastically
in spin and obliquity for the first $3.7\myr$, but then abruptly begins
toggling between two neighboring configurations, one
generating a positive, and the other a negative, component of YORP torque
along the spin axis. The resulting increases and decreases in spin rate
trigger alternating movements of material that convert one configuration
into the other. We invariably see self-governing YORP resulting in strong
self-limitation of spin and obliquity between narrow limits.

Owing to the unavoidable low-amplitude bouncing of spheres
in the HSDEM algorithm, successive appearances of the two configurations are
not quite identical, and so the toggling is not quite periodic. In some
cases we see self-governing come to an end and return to stochastic
evolution, possibly due to this non-repeatability. We can
conjecture that a different computational approach that allows the particles
to come to rest with respect to each other might show truly periodic
switching that continues indefinitely.

Examples of the positive- and negative-torque configurations from simulation
12 are shown in Fig.~\ref{fig:sim12selfgovshapes}, and the torques that
they generate as functions of obliquity are shown in
Fig.~\ref{fig:sim12selfgovtorques}. At obliquity values between
20 and 30 degrees,
where the toggling occurs, one configuration has positive values of the
spin and obliquity torques while the other one has negative values.
A subtle bending at the constriction, one third of the way up from the
bottom, results in a change in sign of the obliquity and the spin torques.
Note that, as a result of the centrifugal kneading, parts of the object
have settled into ordered packing, giving it a ``head-tail'' structure
composed of two more rigid (packed) chunks joined by a flexible waist.
This suggests the possibility that known ``head-tail'' or contact binary
objects might also be found in self-governing states.

\subsubsection{Stagnating YORP}

A close look at Fig.~\ref{fig:newfig8_sim10} shows that
the object in simulation 7 does not reach, or even approach, the expected
YORP-cycle end state of $90\arcdeg$ obliquity and zero spin, but instead
asymptotes to a moderately slow spin rate ($\sim 2$ day period) at an
obliquity of $80\arcdeg$. This is an example of stagnating YORP, which we
see in 4 of the 16 objects simulated at the standard {\tt TACO} resolution
and in 2 of the 4 objects re-run at higher resolution. Technically, stagnation
is just a special case of stochasticity, in which, as a result of
multiple mass movements, an object randomly falls into a configuration of very
low torque. What makes it distinct from stochastic YORP is that objects can
remain ``stuck'' in such configurations for times that approach expected NEA
lifetimes, effectively shutting off their YORP evolution.

\subsubsection{Modified YORP Cycle}\label{s.modyc}

Not every object is equally susceptible to small changes in topography,
and not every shape reconfiguration necessarily reverses the direction of
spin or obliquity evolution. For roughly one-third of our test objects,
shape changes affect only the rate of evolution, and as a result these objects
follow what we refer to as a modified YORP
Cycle. Figure \ref{fig:newfig_sim7} shows simulation 5 as an example.
This object approaches the same end state as its rigid counterpart, in a
shorter elapsed time, having accelerated to, and decelerated from, a lower
maximum spin rate. Another example is seen in the obliquity evolution of
simulation 7 (Fig.~\ref{fig:newfig8_sim10}, lower panel), which is
monotonic and resembles the rigid-body prediction until it stagnates.
Table \ref{tab:spinrates} shows that the duration of the
modified YORP cycle can be either longer or shorter than the rigid-body
cycle.

\citet{Roz13} and \citet{Kaa13} have argued that greater topographic sensitivity
is a characteristic of objects with weaker overall YORP torques, suggesting
that objects that are more instrinsically ``yorpy'' might be more likely
to follow a modified YORP cycle. However, we see no tendency for self-limited
or modified YORP cycle behavior to be correlated with the magnitudes of
either the initial torques on the test objects or the episodic torques during
the aggregate evolution.

\subsubsection{End States}

The fifth column of Table \ref{tab:spinrates} and the third column of
Table \ref{tab:obliquity} give the YORP cycle end-state spins and obliquities,
for objects evolving as rigid bodies. The ninth and fourth columns
(respectively) of those tables give the corresponding quantities for the
aggregate objects at the ends of the simulations.

Two of the 16 objects have rigid-body end states of formally
infinite spin; the remaining 14 have rigid-body end state spins of zero,
reached in finite time $t_{\rm YC}$. As we have emphasized, most
aggregates do not reach or approach the rigid-body end states: only 5
aggregates have spun down to zero or are monotonically decelerating at slow
spin rates at the end of the simulations. One object has fissioned, but the
majority have either stagnated (2) or are stochastically wandering (8) at
finite spin rates, at simulation end times averaging $2.5t_{\rm YC}$.
The situation is similar for obliquity.
All 16 objects have rigid-body obliquity end states at, or nearly at,
$90\arcdeg$. Among the aggregates, roughly half (7) have reached this
obliquity or are clearly on their way there as of the end of the simulation.
Of the 8 remaining objects that do not fission, 3 have reached different
constant values of obliquity, and 5 are wandering stochastically.

Two of the four simulations with higher resolution tilings have rigid-body
end states of formally infinite spin and the other two have rigid-body end
state spins of zero, reached in finite $t_{\rm YC}$ while the four aggregate
objects wander up and down in spin stochastically. In the case of obliquity,
the rigid bodies have end states at obliquity values between $73\arcdeg$ and
$86\arcdeg$ while the aggregates are stochastically wandering. Three of the
four aggregates stagnate at certain values as well, two of them at nearly
$0\arcdeg$.

The clear tendency for a majority of aggregates not to evolve to the standard
YORP-cycle end states has important implications for orbital evolution due
to the Yarkovsky effect. We return to these issues in section
\ref{s.discussion} below.

\subsection{Mass Movement}

An important aspect of the mass motion events is that the moving material is
not restricted to the surface of the object. Animations of the shape
evolution clearly show the entire object reconfiguring (albeit often subtly)
rather than material migrating along the surface. To quantify the amount of
deep motion in each event, we sort the {\tt pkdgrav} spheres in order of
effective potential $\Phi_{\rm eff} = \Phi - {1 \over 2} \omega^2 R^2$,
where $\Phi$ is the gravitational potential, $\omega$ is the rotational angular
frequency, and $R$ is the cylindrical radius from the spin axis. Each sphere
is given an enclosed-mass-fraction coordinate equal to its position in
the sorted list divided by the number of spheres. We then tally the number
of times that the sphere at each mass fraction coordinate moves by
more than $25\%$ of its radius during the simulation.

\begin{figure}[t]
\centerline{ \includegraphics[scale=.55]{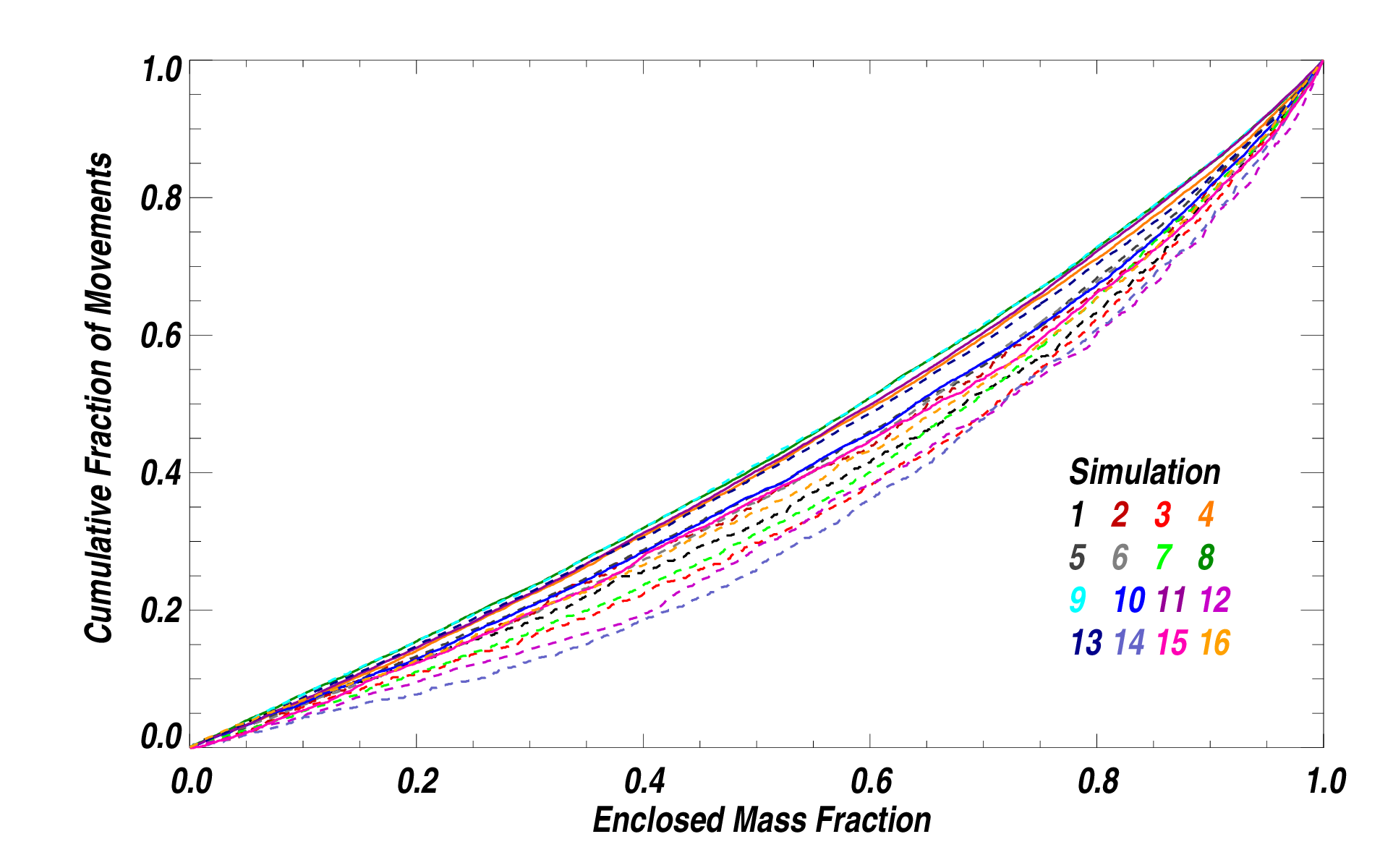} }
\caption{Normalized cumulative distribution of mass-movement events for the
16 simulations as a function of enclosed (by surfaces of constant effective
potential) mass fraction. Each color corresponds to a different simulation.
Solid lines indicate the 5 objects that exhibit mass-loss episodes.
\label{fig:massfrac}
}
\end{figure}

Figure \ref{fig:massfrac} shows the cumulative distributions of the movements
as functions of the mass fraction for all of the simulations. Though the outer
layers are somewhat more mobile, there is clearly
motion of material all the way into the deep interior. Between 25\% and
40\% of the mass motion events occur in the inner (i.e., most tightly bound)
half of the mass. The outer 10\% of the mass accounts for only 15\% to
25\% of the motion. The freedom to reconfigure internally is what gives the
objects the ability to acquire greater rigidity with time, in these
simulations by falling into ordered packing. The figure also shows that
objects that shed mass (solid
curves) tend to exhibit more deep motion than those that do not (dashed curves).
We will see below that this is likely related to large-scale shape changes
that promote mass shedding.

\subsection{Mass Loss and Binary Formation}

\begin{figure}[t]
\centerline{ \includegraphics[scale=0.5]{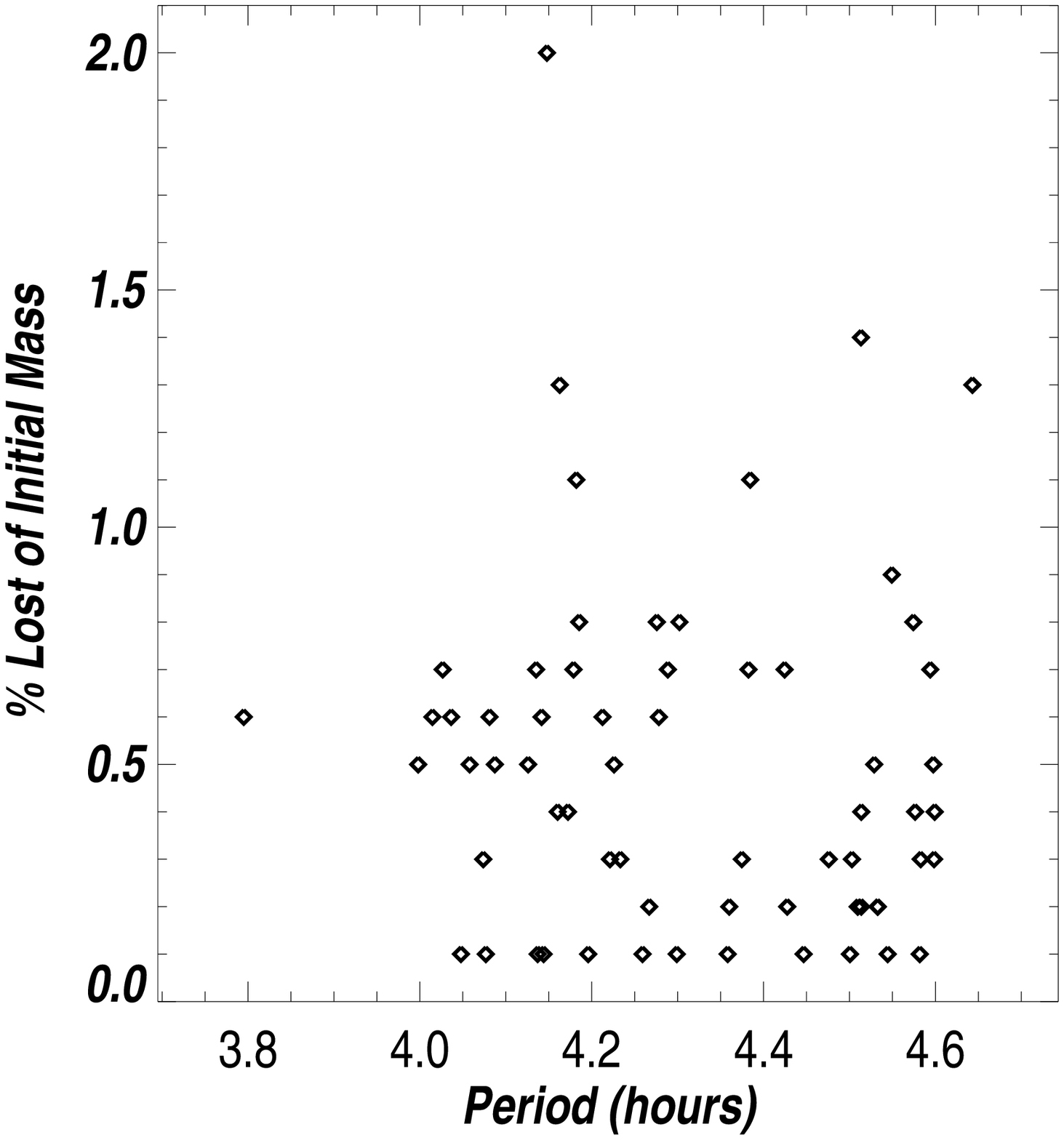} }
\caption{
The distribution of all mass-loss events. The fraction of the initial
body mass that is lost is plotted against the spin period at the time of
mass loss. A single sphere corresponds to $0.1\%$ of the initial mass. Mass
loss occurs at a mean rotation period of 4.3 hours.
\label{fig:events}
}
\end{figure}

Five of the 16 simulations experience mass-loss episodes. The first 4
columns of Table \ref{tab:iso} show the number of mass-loss events, the total
percentage of mass lost from the initial object, and the average time between
events for the 5 simulations. Figure \ref{fig:events} shows the distribution of events in terms of the mass lost per event and the rotation period at the
time of the event. The minimum mass loss in a single event is $0.1\%$ (one
sphere) while the maximum is $2.0\%$. The mass-loss episodes can occur as
isolated events or as a chain of events. The average time between events can be
as short as $0.01\myr$ for consecutive events and more than $1\myr$ for isolated
events.

\begin{table}[t]
{\centering
\caption{Mass-Loss Events}
\label{tab:iso}
\begin{tabular}{*9c}
\hline
Simulation & $N_{\rm ev}$$^{\rm a}$ & $\Delta M$(\%)$^{\rm b}$ &
$\langle \Delta t \rangle$$^{\rm c}$ &
$\left\langle \rho \right\rangle$$^{\rm d}$ &
$P_{\rm min,sph}$$^{\rm e}$ &
$\langle \left({b \over a}\right)_{\rm ev} \rangle$$^{\rm f}$ &
$P_{\rm min,pro}$$^{\rm g}$ &
$\langle P_{\rm ev} \rangle$$^{\rm h}$ \\
\hline
 4 & 10 & 7.0 & 0.05 & 1.42 & 2.77 & 0.76 & 3.18 & 4.22\\
 8 & 10 & 6.1 & 0.01 & 1.53 & 2.67 & 0.52 & 3.72 & 4.49\\
10 & 13 & 6.3 & 0.15 & 1.46 & 2.73 & 0.91 & 2.86 & 4.19\\
11 & 18 & 7.6 & 1.71 & 1.48 & 2.71 & 0.58 & 3.57 & 4.53\\
15 & 13 & 5.1 & 0.04 & 1.57 & 2.63 & 0.71 & 3.12 & 4.08\\
\hline\\
\end{tabular}\par}
$^{\rm a}$ Number of events.
$^{\rm b}$ Total mass lost (percent).
$^{\rm c}$ Mean time between events (Myr).
$^{\rm d}$ Mean bulk density at time of mass loss ($\g\cm^{-3}$).
$^{\rm e}$ Minimum spin period (h) for cohesionless sphere of the same density.
$^{\rm f}$ Mean axis ratio at time of mass loss.
$^{\rm g}$ Minimum spin period (h) for cohesionless prolate spheroid of the same density and axis ratio.
$^{\rm h}$ Mean spin period (h) at time of mass loss.
\end{table}

\begin{figure}[t]
\centerline{ \includegraphics[scale=0.5]{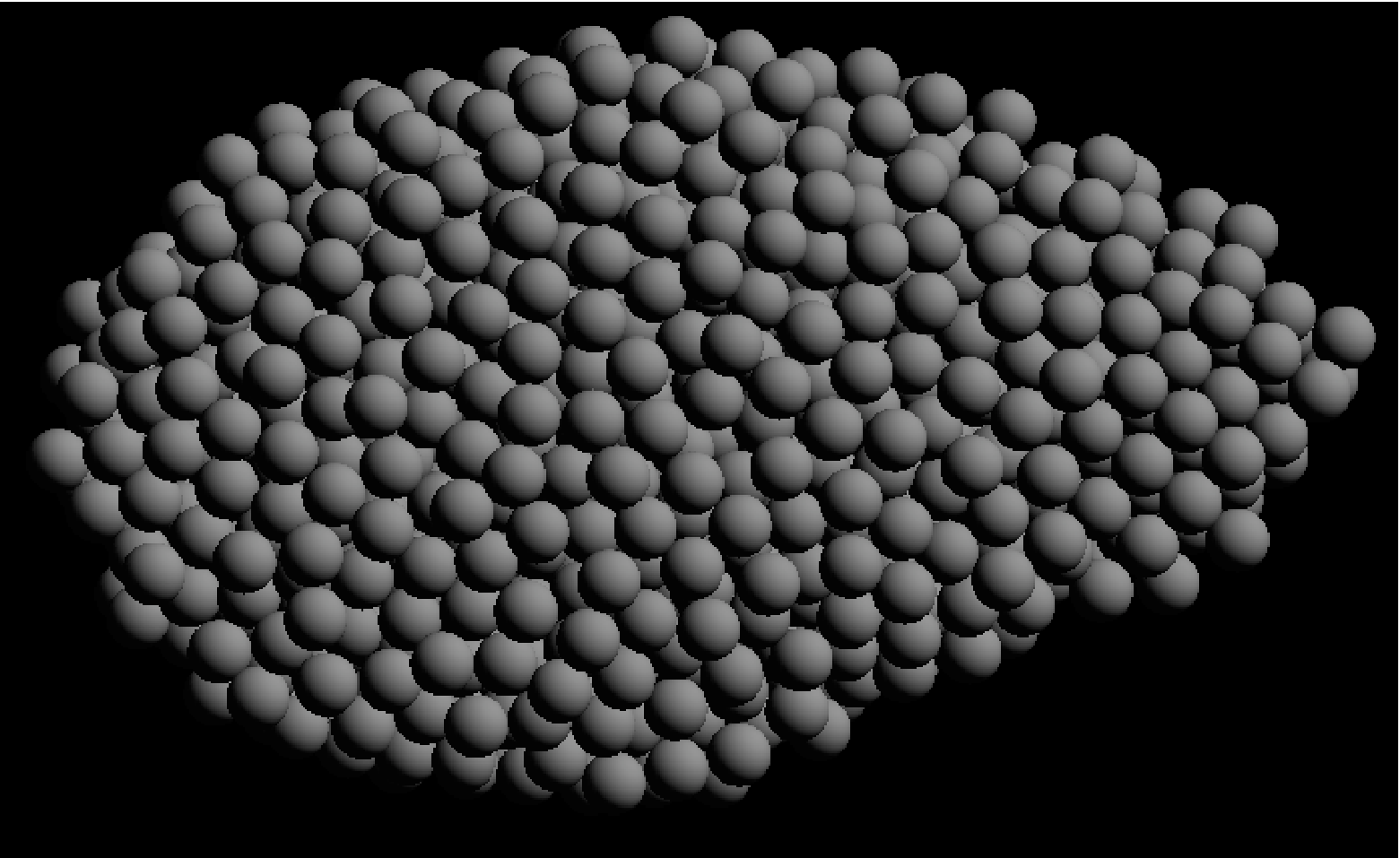} }
\caption{
The shape of the aggregate object in simulation 11
before having its first mass-loss episode, in which it loses 3 spheres. The
view is along the rotation axis. Material is lost from the right side.
\label{fig:aggobsim15}
}
\end{figure}

The rotation periods at which mass loss occurs range from 3.8 to 4.6 hours, with
a mean of 4.3 hours. The averages for each simulation are listed in the last
column of Table \ref{tab:iso}. These spin rates are substantially slower than
the nominal spin limits at which loose material should become unbound from
the equator of a sphere with the same bulk density. The fifth and sixth columns
of Table \ref{tab:iso} give the densities for each simulation, averaged over
mass-loss events, and the corresponding limiting periods for spheres. The
latter are between 2 and 3 hours. Part of the difference can be attributed
to the fact that in our simulations mass is commonly lost from one end of the
object as its shape becomes elongated (an example, just before the event,
is shown in Fig.~\ref{fig:aggobsim15}). The axis ratios $b/a$ in the plane
normal to the spin axis, again averaged over events for each simulation,
are given in column 7 of Table \ref{tab:iso}, and column 8 gives the limiting
spin period for prolate spheroids of the same axis ratio and density
\citep{Har96,Ric05}.
The theoretical limits are still 20\% to 30\% slower than the simulation
results. We can speculate that this difference may be caused by the tendency
for our objects to become sharply pointed at the ends, by non-uniformity
of the interior bulk density,
or by the dynamical motion of the material close to the tip.

\begin{figure}[t]
\centerline{ \includegraphics[scale=0.55]{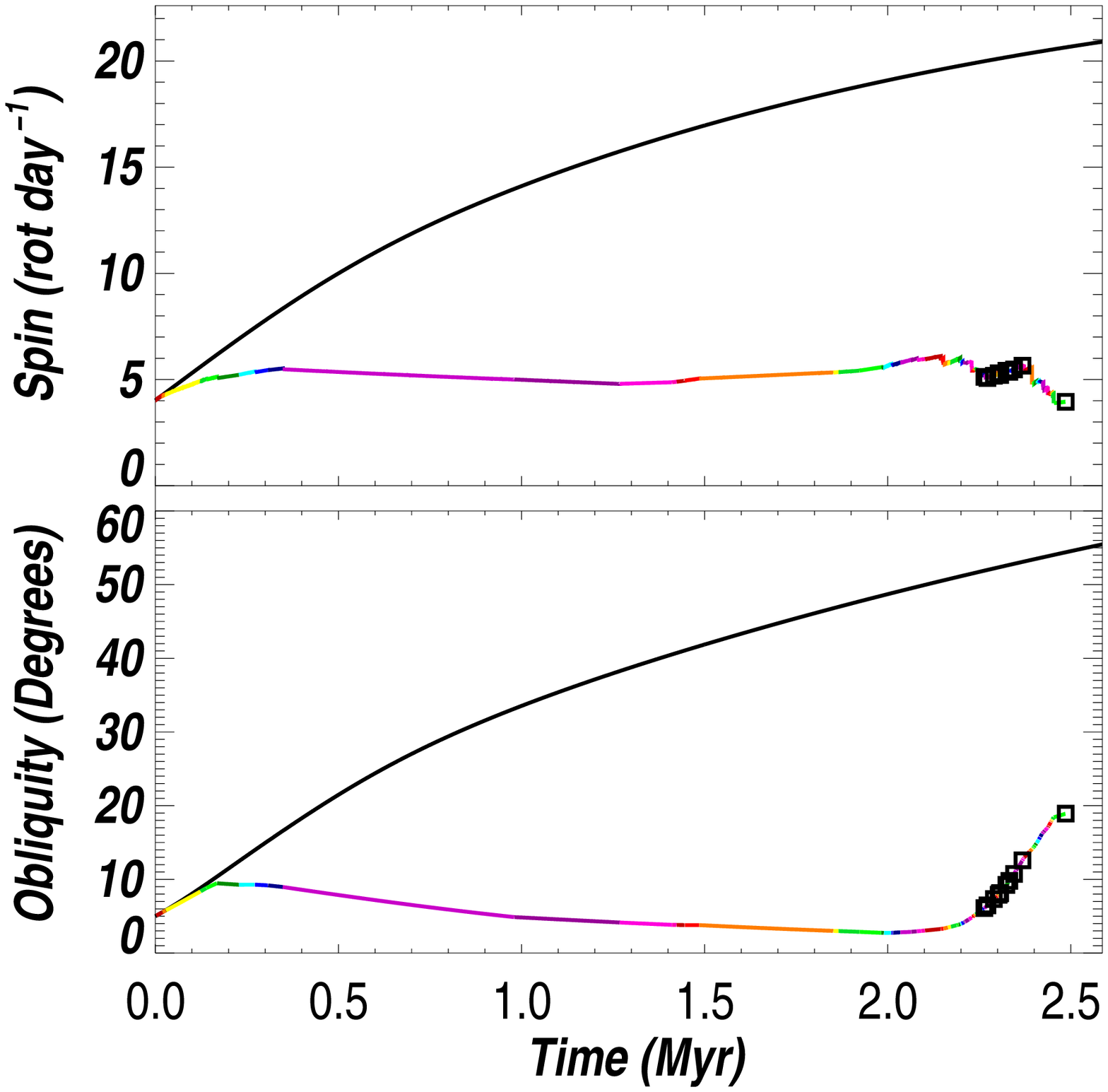} }
\caption{
The spin ({\em top\/}) and obliquity ({\em bottom\/}) evolutions of the
object in simulation 8, as in Fig.~\protect{\ref{fig:newfig8_sim10}}.
{\em Black\/} and {\em colored lines\/} show the rigid-body evolution and
aggregate evolution, respectively. {\em Black squares\/} indicate mass-loss
episodes. At the end of the simulation the object fissions to form a binary.
\label{fig:binaryevo}
}
\end{figure}

\begin{figure}[t]
\centerline{ \includegraphics[scale=0.5]{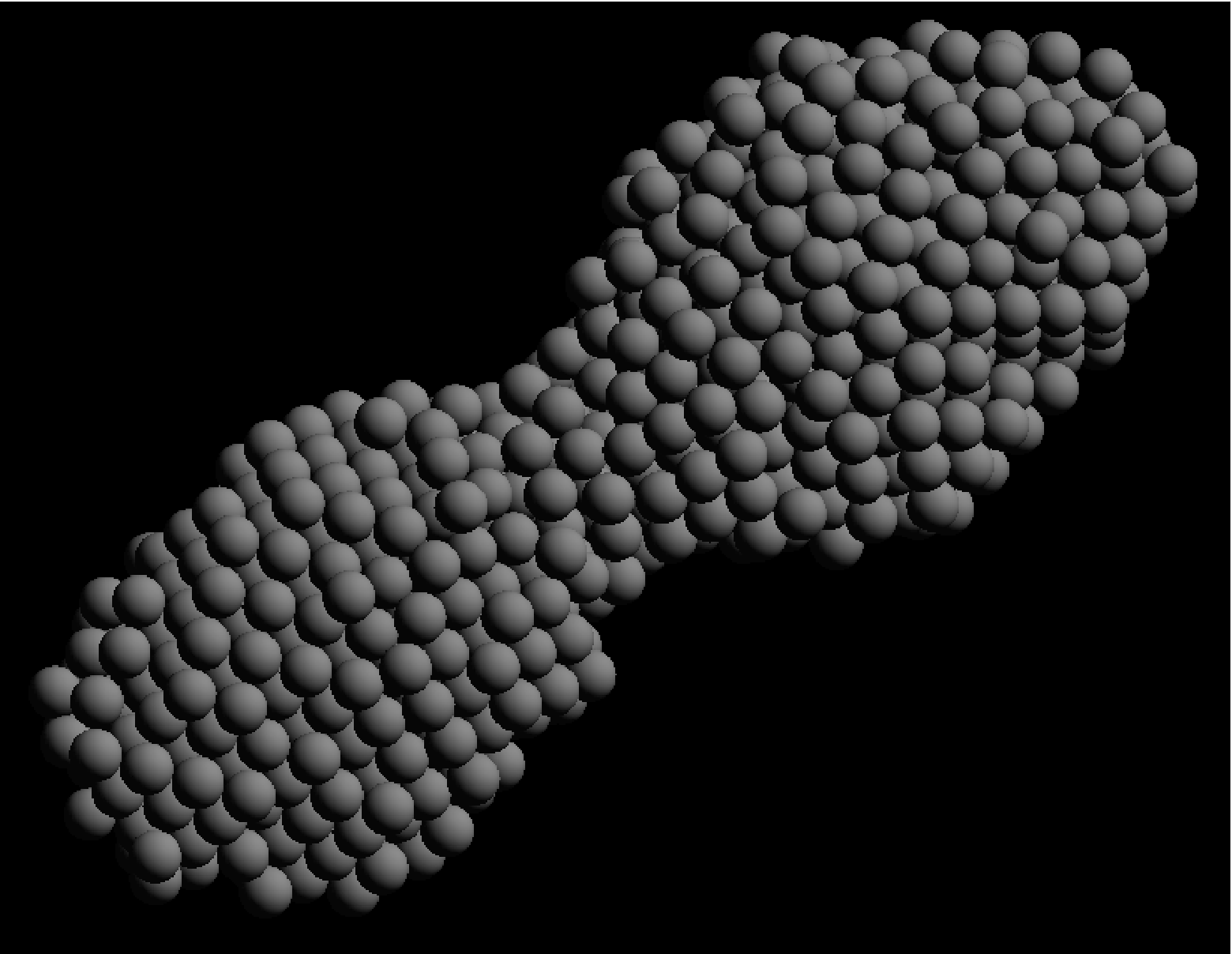} }
\caption{
The shape of the aggregate in simulation 8 at the last point of
contact before splitting and forming a binary asteroid. The view is along the
rotation axis.
\label{fig:bilast}
}
\end{figure}

The spheres are removed from the simulations after being shed from the main
object. We do not track their orbital evolution since the main objects become
strongly prolate. \citet{Sch07b} has shown that objects orbiting a rapidly
rotating prolate body would most likely escape rather than reach stable orbits
where they could accrete to form a binary companion. However, we do encounter
one case of binary formation. Figure \ref{fig:binaryevo} shows
the spin and obliquity evolutions of the object in simulation 8. Black squares
indicate mass loss episodes. After losing $6.1\%$ of its initial mass in 9
events, the objects splits in two (at a time of about 2.5 Myr, at which point
the simulation is stopped). At the moment of fission, the object is increasing
in angular momentum but decreasing in spin rate because of its evolution
toward an elongated shape. Figure \ref{fig:bilast} shows the aggregate at
the last point of contact. Note the wasp-waist constriction, where the fission
occurs. After fission, the primary object contains $52.7\%$ of the initial
mass while the secondary contains $41.2\%$.

\subsection{Axis-Ratio Evolution}

\begin{figure}[t]
\centerline{ \includegraphics[scale=0.5]{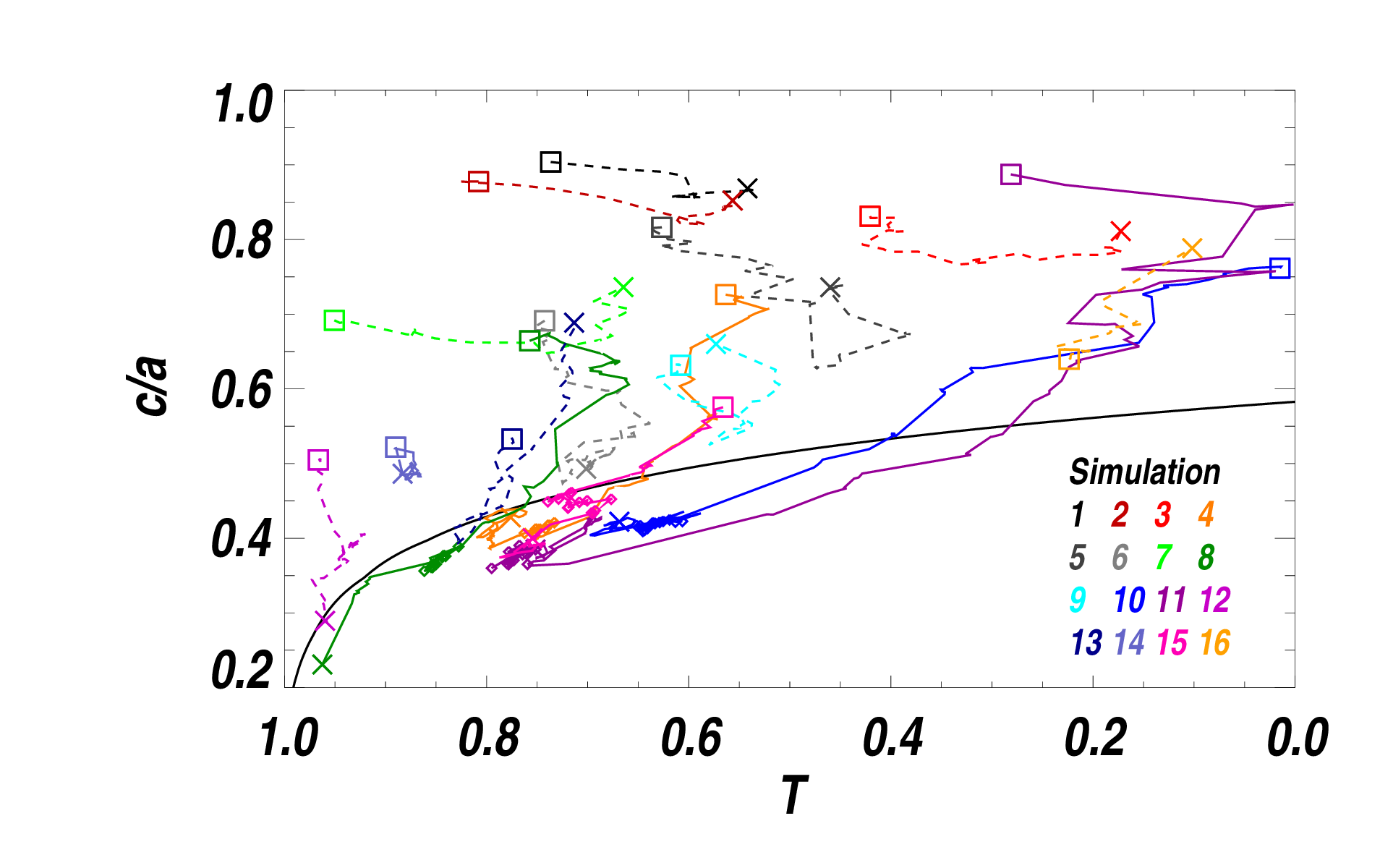} }
\caption{
Shape evolution of simulated aggregates, in terms of triaxiality
$T$ and semi-axis ratio $c/a$, as in Fig.~\ref{fig:initialdisorig}.
Each color corresponds to a different simulation. {\em Squares\/} and
{\em crosses\/} indicate shapes at the start and end of each simulation,
respectively. The 5 simulations that shed mass are plotted with
{\em solid lines}; mass-loss epsiodes are marked with {\em small diamonds}.
The smooth black curve indicates the sequence of Jacobi ellipsoids.
\label{fig:axisandtevo}
}
\end{figure}

Figure \ref{fig:axisandtevo} shows the evolution of the 16 objects in the
space of triaxiality $T$ and semi-axis ratio $c/a$, where $T$
is given by equation \ref{eq:triax}. Each simulation is shown in a
different color. Squares indicate the initial objects, as in
Fig.~\ref{fig:initialdisorig}; each object at the end of the simulation is
indicated by an X. Solid curves identify the 5 objects that lose mass.
Most objects are still evolving in spin at the end of the
simulation and therefore the X does not necessarily represent an evolutionary
end point.

The majority of objects (13 of 16) become flatter ($c/a$ decreases) during
the simulation. Evolution in triaxiality can go in either direction, but
we note two striking trends. First, all objects following a modified
YORP cycle (simulations 1, 2, 3, 5, and 16) evolve toward smaller $T$; that is,
they become less prolate and more oblate. This appears to be a result
of the deceleration to very slow spin rates, although it is important to
note that the evolution does not follow a fluid sequence, which would
imply $T \to 0$ at finite $\omega$. These objects arrive at genuine non-rotating
end states with non-zero $T$. Second, all of the objects that shed mass
or fission (simulations 4, 8, 10, 11, and 15) evolve toward smaller $c/a$
and larger $T$; that is, they become highly elongated and prolate. Moreover,
these 5 simulations show the largest changes in axis ratio. We 
conjecture that these objects were the most initially deformable, which is
consistent with the finding (Fig.~\ref{fig:massfrac}) that they also show
the greatest amount of deep mass motion. The smooth black line in
Fig.~\ref{fig:axisandtevo} indicates the sequence of Jacobi ellipsoids.
The objects
that lose mass are the only objects that dip well below the Jacobi sequence,
and the episodes of mass loss from the endpoints (small diamonds) occur
exclusively below the sequence, as does the final fissioning of simulation
8.\footnote{It is interesting that the only other object that
becomes as elongated as the fission case, simulation 8 (Fig.~\ref{fig:bilast}),
is simulation 12 (Fig.~\ref{fig:sim12selfgovshapes}),
which evidently escapes fission by self-governing.}

\subsection{The Statistical Spin and Obliquity Evolutions}

Stochastic processes, while non-deterministic, can still be described
statistically. Here we formulate
a statistical description of the spin-state evolution obtained from our
simulations, intended to inform models of spin-dependent processes,
particularly the Yarkovsky effect. This description should be regarded as very
preliminary: first, because our initial conditions were chosen to survey a
variety of interesting YORP behaviors, and not to represent a realistic
population of objects; and second, because at this point we are still
neglecting processes known to be important, such as thermal conduction.

We break the evolution into discrete intervals bounded by material movements
(shape changes). The shape is constant (except for small bouncing of the
spheres) during each interval. Consider one such interval of duration $\delta
t$, over which the change in spin rate is $\delta \omega$. We define the
statistical spin evolution, $\alpha_s$, by
\begin{equation}
\alpha_s = \pm
\left| {\delta \omega} \over {\delta t} \right|,
\end{equation}\label{e.alphas}
where the sign
is determined by whether the sign of $\delta \omega$ is the same as ($+$) or
opposite to ($-$) that in the previous interval. With this definition, rapidly
alternating spin-up and spin-down behavior, characteristic of self-governed
YORP, would be described by consistently negative values of $\alpha_s$. Strongly
stochastic YORP would produce a tendency for negative $\alpha_s$, while
weak stochasticity or modified YORP cycle evolution would appear as
predominantly positive values.
 
Similarly, we define the statistical obliquity evolution, $\zeta_s$, in terms of
the change of obliquity during the interval according to
\begin{equation}
\zeta_s = \pm \left| {\delta \epsilon} \over {\delta t} \right|,
\end{equation}\label{e.zetas}
with the sign determined by comparison with the previous interval, as above.
 
\begin{figure}[t]
\centerline{ \includegraphics[scale=0.6]{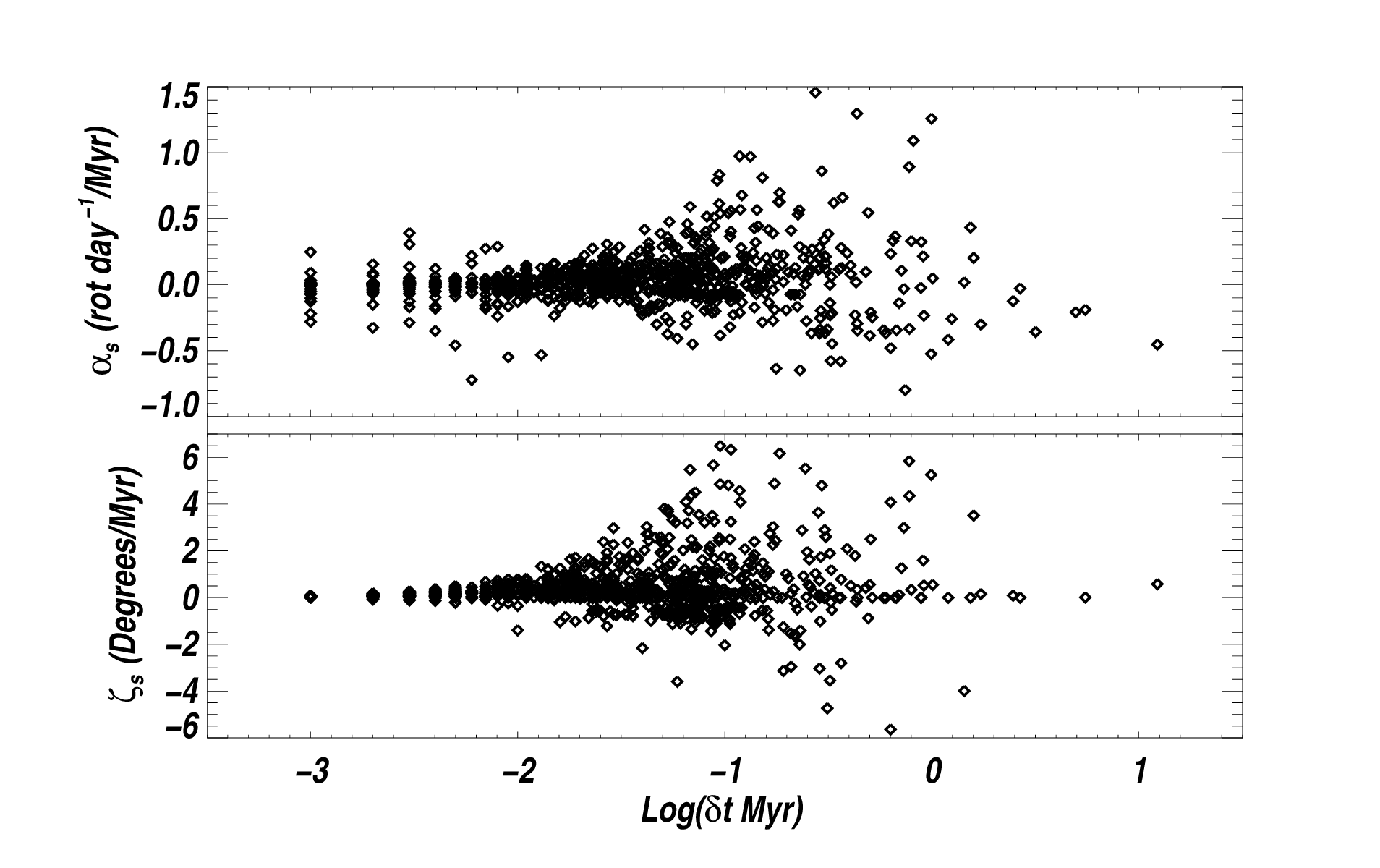} }
\caption{Scatter plots of the distribution of ({\em top\/})
$(\alpha_{s},\delta t)$,  and ({\em bottom\/}) $(\zeta_s, \delta t)$,
defined in Eqs.~(5) and (6).
The horizontal axis is the base-10 logarithm of the time interval $\delta t$ in Myr.
\label{fig:scatter}
}
\end{figure}

Figure \ref{fig:scatter} shows the joint distribution in $(\alpha_s, \delta
t)$ and $(\zeta_s, \delta t)$ for all intervals in all simulations. One can
see that typically a few $10^4\yr$ elapses between shape changes, and
significant alterations to the spin evolution can generally be expected
on $\sim10^5\yr$ time scales, for the km-sized objects considered here.
The shapes of the distributions in $\alpha_s$ and $\zeta_s$ are actually
surprisingly similar, showing a slight overall tendency toward weak
stochasticity. Similar $(\alpha_s,\zeta_s,\delta t)$ distributions derived
from unbiased initial conditions, with all relevant input physics, will
provide a pathway for stochastic YORP to be included, in a Monte Carlo
sense, in simulations of orbit evolution that incude the Yarkovsky effect.

\section{Discussion}\label{s.discussion}

The simulations presented here strongly support the conjecture of
\citet{StatlerYORP} that YORP can behave stochastically when the surface
topography is susceptible to spin-driven alterations. We actually see four
distinct types of behavior, three of which---stochastic YORP, self-governed
YORP, and stagnating YORP---collectively give rise to the phenomenon of
YORP self-limitation.

It is a widely held view that YORP is responsible for the formation
of top-shaped asteroids, and particularly top-shaped asteroids with binary
companions. This view has been shaped, in large part, by the
influential simulations of \citet{WalshNature}. By continually adding
angular momentum, ostensibly supplied by YORP, the authors were able to
make idealized aggregates evolve, through motion of surface
material, to nearly axisymmetric top shapes with equatorial ridges, which
then shed mass that accreted in orbit to form binaries.
But as we have demonstrated in this paper, YORP should not be presumed to
be an inexhaustable source of angular momentum. Self-limitation is likely
to intervene, possibly stalling the mechanism before substantial evolution has
a chance to occur.

In a follow-up study, \citet{Wal12} show that the distinctive evolutionary
path taken by their earlier aggregates was in part a consequence of the
rigidity resulting from the initial HCP arrangement
of identical spheres. Altering the size distribution or the initial
arrangement to avoid HCP makes the aggregates more fluid, and tends to
inhibit both evolution to axisymmetry and mass shedding. Our test objects,
while composed of identical spheres, start with a disordered packing.
Except for those that follow a modified YORP cycle, our objects generally
evolve toward more elongated shapes. The minority that do lose mass do so
mostly in modest amounts from their elongated tips; we see only one case of
fission into two comparably sized bodies. In these respects their behavior
falls between the ``near-fluid'' and ``intermediate'' cases of \citet{Wal12}.
Owing to the strongly time-varying potential, we would not expect the
slowly-shed mass to accumulate on stable orbits, and consequently not
form long-lived binaries.

Exactly how YORP self-limitation may occur on objects with less deformable
interiors is a question that we cannot fully address with the present set of
simulations. We do observe parts of our aggregates, through time-varying
centrifugal massaging, occasionally falling into a HCP state. This tends
to happen after some amount of reshaping has already occurred, leading to
localized off-center chunks of higher rigidity rather than a central
rigid core. We would not expect real objects to ``crystallize'' in this way,
but a real aggregate may naturally develop rigidity as a result of
time varying stresses that allow its components to find interlocking
configurations. This is a question for future simulations that can take
more physical effects into account with greater realism (see below).

In approximately two-thirds of the cases we have simulated, YORP
self-limitation prevents objects from decelerating to
zero spin rate. These objects never complete a YORP cycle; consequently
one would expect them to avoid a chaotic tumbling phase, and hence to
preserve their original sense of spin (direct or retrograde). This has
bearing on the fraction of retrograde rotators among the NEAs.  Studies of
the delivery of NEAs from the Main Belt \citep{Bot02, LaS04} conclude that
approximately $37\%$ should arrive as a result of inward Yarkovsky drift,
requiring retrograde rotation, into the $\nu_{6}$ secular resonance. The
rest should come through other resonances with an equal fraction of retrograde
and direct rotators. We should therefore expect about $69\%$ of NEAs to have
been {\em delivered\/} to their current orbits with retrograde rotation.
But from measurements of Yarkovsky drift in the {\em present\/} NEA
population, \citet{Far13} estimate the retrograde fraction to be just that,
i.e., $69\%$, implying that few of the objects have forgotten their
original spin sense. Self-limitation may possibly account for this. If we
assume that all NEAs with Yarkovsky drift measurements are aggregates, then
about $25\%$ out of the $37\%$ of NEAs delivered through $\nu_{6}$ will be
prevented from forgetting their initial senses of spin. Adding half of the
$12\%$ of the $\nu_{6}$ objects that do forget, plus half of the $63\%$ that
come through other resonances, we should expect roughly $63\%$ retrograde
rotators, not inconsistent with the observational result. While the
statistical uncertainties are substantial, and our simulations are not yet
definitive, YORP self-limitation may provide a means to reconcile the high
present-day retrograde fraction with the long lifetimes of NEAs relative to
their nominal YORP-cycle timescales.

The tendency for self-limited, and particularly stochastic, YORP to preserve
a memory of earlier spin states is also relevant to the spreading of
collisional families by the Yarkovsky effect. \citet{Bot13DPS} find that
the envelopes, in $(a,H)$ space, of old families (ages $\sim 1\gyr$)
are inadequately fit by models in which the spin sense of objects
is frequently reset, as would happen at the end of a YORP cycle. Instead,
a stochastic YORP model, in which the memory of the spin state, and hence the
direction of Yarkovsky drift, is preserved for longer times, results in a
much better fit. These results are encouraging for the general picture of
stochastic YORP, and furthermore hint that even relatively small collisional
fragments in the Main Belt may be re-accreted aggregates. We can anticipate
that a statistical description, as in Fig.~\ref{fig:scatter}, of future
results of a more exhaustive, unbiased suite of simulations will help
to clarify the situation further.

In the interest of computational expediency, we
have neglected physical effects that are known to be important to YORP,
and therefore the simulations presented here should be interpreted as
a first demonstration of processes that {\em may\/} occur, and not (yet) a
definitive depiction of what {\em does\/} occur. The key effects to be
explored in future simulations should include:
\begin{itemize}
\item
{\em Thermal conduction\/}: At a given orientation, this has no direct
impact on the spin component of torque, but does affect the obliquity
component. Since all components are obliquity-dependent, the coupled
evolution will change. One indicator of this dependence is that 
the rigid-body, YORP-cycle obliquity end states are expected to be concentrated
near $0\arcdeg$ for direct rotators with moderate thermal inertia
$\Gamma$ \citep{Cap04}, rather than $90\arcdeg$ when $\Gamma=0$ (see
Table \ref{tab:obliquity}). As a check, we have computed the torques
on our initial objects, taking into account thermal conduction as well as
self-heating (see below), with an assumed $\Gamma = 200\tiunits$, and
verified that the rigid body end states do, in fact, shift to $0\arcdeg$.
\item
{\em Self-heating\/}: Where concavities exist, parts of the surface can
be heated by light reflected or radiated from other parts of the surface.
This effect tends to reduce local temperature gradients caused by
self-shadowing, which \citet{Roz13} argue may somewhat lessen the sensitivity
of YORP to small surface changes. To gauge the potential influence of this
effect on our results, we have recalculated the YORP curves for the sequence
of objects in simulation 14 (Fig.\ \ref{fig:newfig_torques}) with a full
treatment of self-heating and partial sky blockage. While the torques on
individual objects are changed by typically 10\% to 50\%, the variety and
spread of YORP curves in Fig.\ \ref{fig:newfig_torques} is qualitatively
unaltered. Hence we expect YORP self-limitation and stochasticity
still to occur. \citet{Roz13} further suggest that self-heating will act
to prevent cases in which the spin component of torque has the same sign
at all obliquities. We have ``spot-checked'' this suggestion on a few
objects, including our one initial object that shows a
purely positive spin torque. We do find a tendency for these YORP curves
to be shifted vertically so that they cross zero. This effect may have
bearing on self-governing, which, in our simulations, tends to take
advantage of these single-sign configurations (e.g.,
Fig.\ \ref{fig:sim12selfgovtorques}). However, not all of our self-governing
objects rely on such configurations; and furthermore, eliminating the
single-sign cases does not preclude the possibility of self-governing at
a different obliquity, or of the object finding a different nearby pair of
configurations that are self-governing. Settling the issue of whether
self-heating prevents self-governing will require
calculating the full self-consistent evolution with all relevant thermal
effects included.
\item
{\em Friction and cohesion\/}: The hard-sphere approach to contact
physics is only one of several alternatives, and there are indications
that it may not be the optimal choice for the dense regime in which
particles spend more time in contact than apart \citep{Ric11}. One
recently developed approach is the soft-sphere discrete element method
(SSDEM), newly implemented in {\tt pkdgrav} by \citet{Sch12}. SSDEM permits
a more accurate treatment of multicontact physics, including self-consistent
treatment of sliding and rolling friction and interparticle cohesion.
New numerical experiements on disruptive collisions using SSDEM \citep{Bal14}
are largely in accord with earlier experiments using HSDEM \citep{Lei00},
so one does not expect results for spin-driven
reshaping to differ qualitatively purely because of the computational approach.
However in future work we plan to take advantage of the capabilities of SSDEM
in order to explore a wider range of material properties, to more
realistically account for the effects of irregular particle shapes, and
to test strength models for cohesive aggregates \citep[e.g.,][]{Sch13,San14}.
Recent observational results strongly suggest that cohesive forces are
important both in maintaining the integrity of rapidly rotating objects
\citep{Roz14da} and in influencing the mode of mass loss \citep{Hir14}.
\end{itemize}

\section{Summary}\label{s.summary}

We have presented the first self-consistent simulations of the coupled
spin and shape evolutions of small gravitational aggregates under the
influence of the YORP effect. Because of the sensitivity of YORP to 
detailed surface topography, even small centrifugally driven reconfigurations
of an aggregate can alter the YORP torque dramatically, resulting in spin
evolution that is, in the strong majority of cases, qualitatively different
from the rigid-body prediction.

One-third of the objects simulated follow a simple evolution that can be
described as a {\em modified YORP cycle}. Two-thirds exhibit one or more
of three distinct behaviors---{\em stochastic YORP}, {\em self-governed
YORP}, and {\em stagnating YORP\/}---which together result in {\em YORP
self-limitation}. Self-limitation has the effect of confining the rotation
rates of evolving aggregates to far narrower ranges than would be expected
in the YORP-cycle picture, and greatly prolonging the times over which
objects can preserve their sense of rotation (direct or retrograde).

The simulated asteroids we have tested are initially randomly packed,
disordered aggregates of identical spheres that collectively have a low
internal angle of friction. They are highly deformable and lie near, but not
on the Maclaurin/Jacobi sequence.  Their evolution in shape is charaterized by
rearrangement of the entire body, including the deep interior, and not
predominantly by movement of surface material. Unlike the high-friction-angle
initial configurations tested by \citet{WalshNature}, they do not evolve
to axisymmetric
top shapes with equatorial ridges. When they lose mass, they generally
do so in small amounts from the ends of a prolate-triaxial body, and
always after crossing the Jacobi ellipsoid sequence.

YORP self-limitation may inhibit the formation of top-shapes, binaries, or
both, by restricting the amount of angular momentum that can be imparted to
a deformable body. Stochastic YORP, in particular, will affect the evolution
of collisional families whose orbits drift apart under the influence of
Yarkovsky forces, in observable ways.

\bigskip
The authors are grateful to colleagues for helpful comments
during the course of this work, including David Rubincam, Steve Paddack,
Steve Chesley, Dan Scheeres, Bill Bottke, David Vokrouhlick\'y, and Mangala
Sharma. TSS and DCF were supported in part by NASA Planetary Geology \&
Geophysics grant NNX11AP15G. TSS was also supported by the Independent
Research and Development program while on detail to NSF under the
Intergovernmental Personnel Act; the results and opinions
expressed in this paper are those of the authors and do not reflect the
views of the National Science Foundation. DCF received additional support
from a NASA Harriet G. Jenkins Predoctoral Fellowship and a NASA Ohio Space
Grant Consortium Doctoral Fellowship. DCR acknowledges support from NASA
Planetary Geology \& Geophysics grant NNX08AM39G and National Science
Foundation grant AST1009579.
PT acknowledges the support of the Programme Nationale de Planetologie,
France.
This work has made use of NASA's Astrophysics Data System Bibliographic
Services.

\appendix
\section{Details of the Inter-Code Orchestration}

\begin{figure}[p] 
\epsfig{file=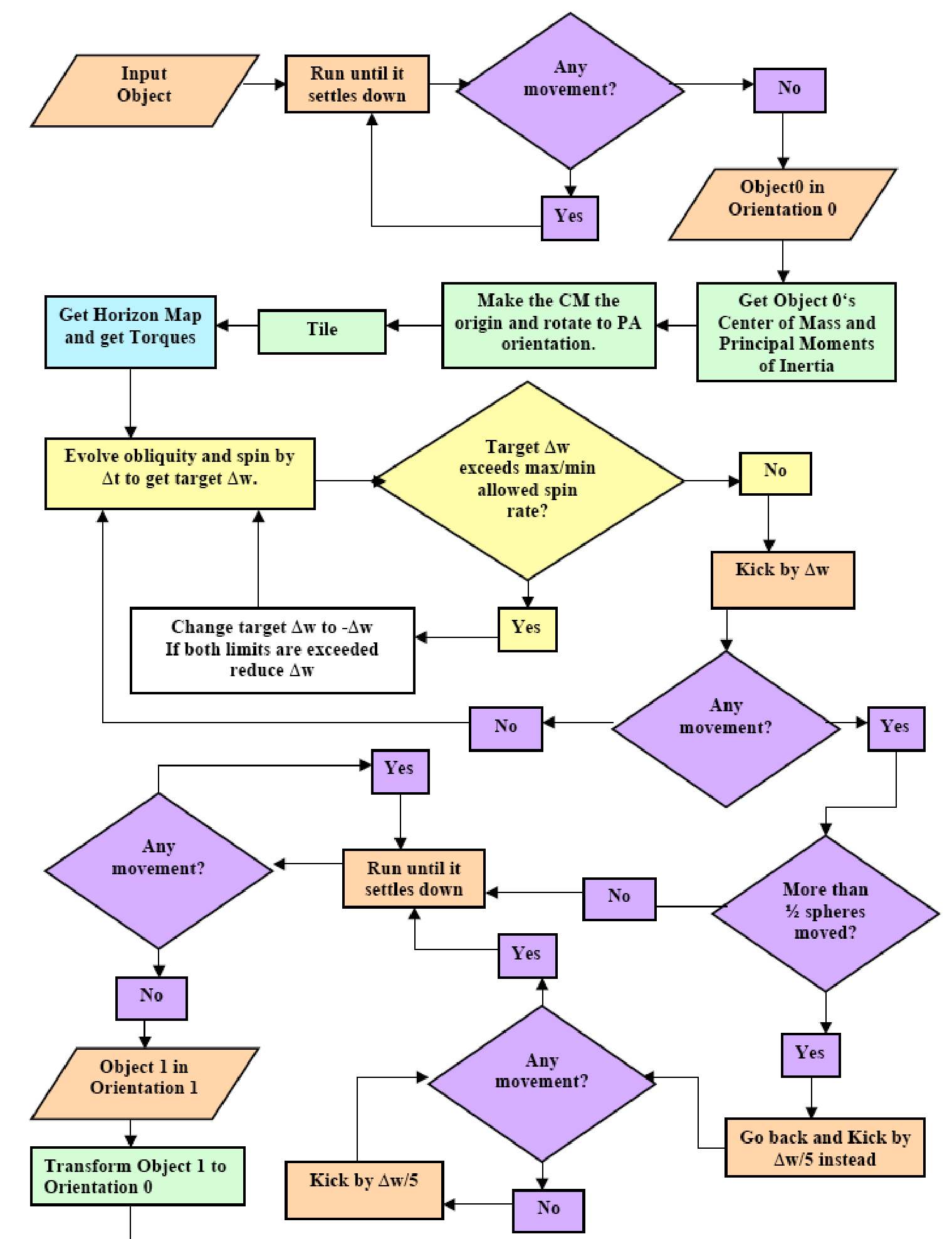, scale=0.6} 
\caption{
Orchestration flowchart; colors indicate major code elements: {\it TACO}
(blue), {\tt pkdgrav} (orange), Spin State Evolution (yellow), Transformation
(purple) and Tiling (green).
\label{fig:flowcharta}
}
\end{figure}

\begin{figure}[p] 
\epsfig{file=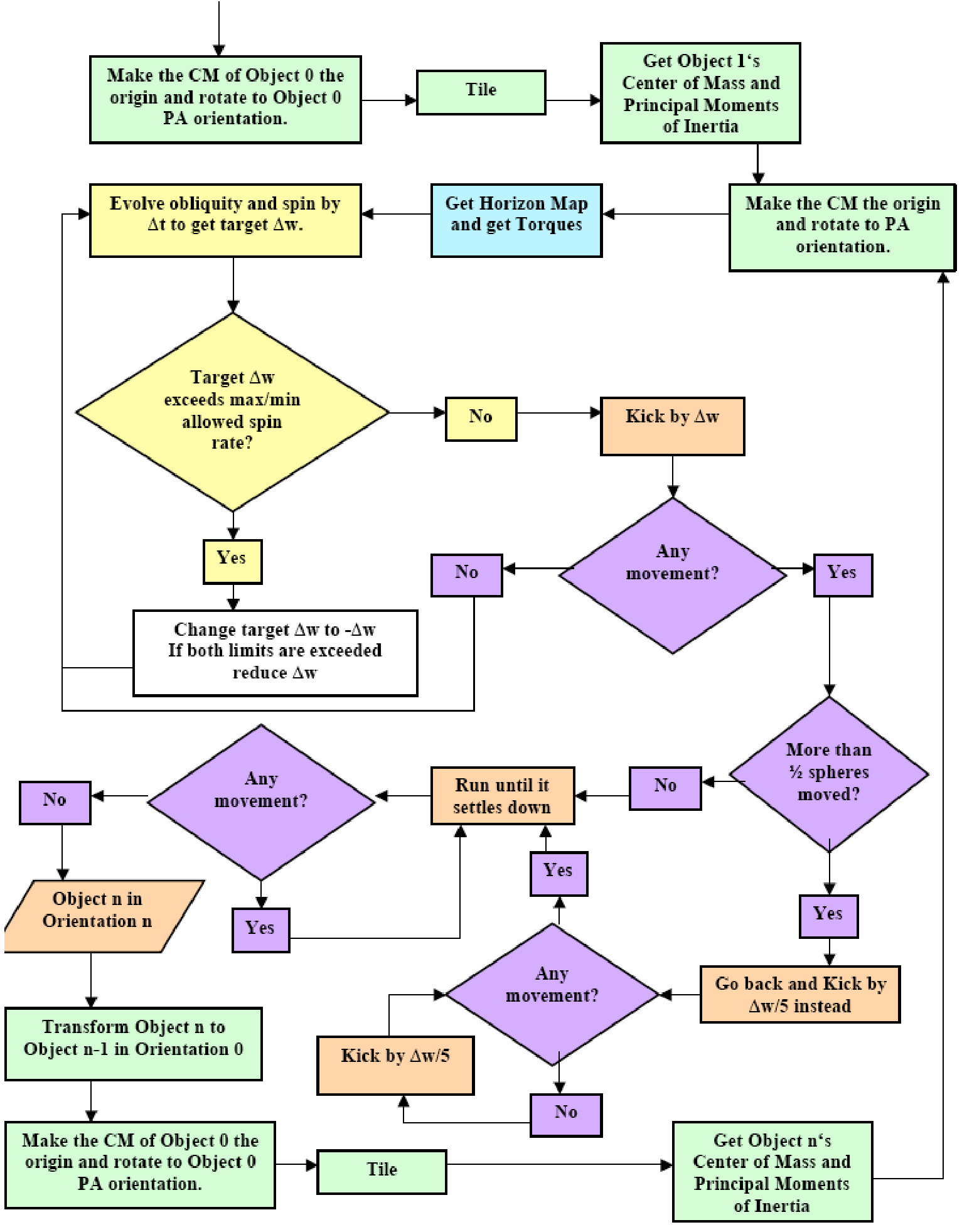, scale=0.6} 
\caption{
Orchestration flowchart, continued.
\label{fig:flowchartb}
}
\end{figure}

Figures \ref{fig:flowcharta} and \ref{fig:flowchartb} show a flowchart of
the full simulation procedure. Orchestration is handled by a {\tt python}
script, which runs and transfers data between the routines of {\tt TACO}
that compute YORP torques (blue), those of {\tt pkdgrav} that integrate the
particle dynamics (orange), and the additional routines that evolve
the spin and obliquity with time (yellow), transform the object in
orientation (purple) and tile the object (green).

We start by dynamically evolving the original object for several rotations at
the initial spin rate using {\tt pkdgrav}. Then we obtain the tiling for the
original object (Object 0) with its center of mass as the origin and in
principal axis orientation. The YORP torques for the Object 0
are then obtained and we evolve the obliquity and spin in time until the spin
rate changes by $0.5\%$ or reaches an extremum. The object is run dynamically
with {\tt pkdgrav} at the new spin rate for several rotations; if there is a
movement of spheres we let it evolve for several more rotations until there
are no more movements. If, at any movement of material, more than half of the
spheres move, we take that as an indication that our increment in spin rate
may have been too large. In that case we go back and increment the spin in
time instead by $1/5$ of the previous increment until a movement occurs. Once
the object has settled down, it is defined as Object 1. We transform it to the
orientation of Object 0 in order to obtain the tiling; this guarantees that
the tiling is altered only over the regions where motion
occurred. Once the tiling is obtained, Object 1 is transformed to its principal
axis orientation with its center of mass as the origin and the YORP torques
are obtained. The obliquity and spin are evolved in time using the torques of
Object 1 until the spin rate changes by $0.5\%$ or reaches an extremum. The
object is run dynamically with {\tt pkdgrav} at the new spin rate for several
rotations until there is another movement and a new object is defined,
repeating the whole process for each new object (Object n) as was done
with Object 1.

\bibliography{asteroids}
\bibliographystyle{elsarticle-harv}

\end{document}